\documentclass[aip,jcp,preprint]{revtex4-1}

\usepackage{amsmath}
\usepackage{amsfonts}
\usepackage{amssymb}
\usepackage{graphicx}
\usepackage{subscript}

\usepackage[utf8x]{inputenc}
\usepackage{color}
\usepackage{bbm}
\usepackage{bm}
\usepackage{float}

\usepackage{comment}

\usepackage{setspace}

\begin{document}

\title{Coarse-graining strategy for molecular pair interactions: A reaction coordinate study for two- and three-dimensional systems}

 \author{Thomas Heinemann}
 \email{thomasheinemann@snu.ac.kr}
 \affiliation{
 Department of Chemistry, Seoul National University, Seoul 08826, Korea  }
\affiliation{Institut f\"ur Theoretische Physik, Technische
  Universit\"at Berlin, Hardenbergstra{\ss}e 36, 10623 Berlin,
  Germany}
\author{Sabine H.L. Klapp}
\email{klapp@physik.tu-berlin.de}
\affiliation{Institut f\"ur Theoretische Physik, Technische
  Universit\"at Berlin, Hardenbergstra{\ss}e 36, 10623 Berlin,
  Germany}

\begin{abstract} 

We investigate and provide optimal sets of reaction coordinates for { mixed pairs of
molecules displaying polar, uniaxial or spherical symmetry in two and three dimensions.
These coordinates are non-redundant, i.e., they implicitly involve the molecules' symmetries.
By tabulating pair interactions in these coordinates, resulting tables are thus minimal in length and require a minimal memory space.
}
The intended fields of application are computer simulations of large ensembles of molecules or colloids { with rather complex interactions} in a fluid or liquid crystalline phase { at low densities.}
Using effective interactions directly in form of tables can help bridging the time and length scales without introducing errors stemming from any modeling procedure.
Finally, we outline { an exemplary} computational methodology for gaining an effective pair potential { in these coordinates}, based on the Boltzmann inversion principle, by providing a step-by-step recipe.

\end{abstract}

\maketitle

\section{Introduction}

Simulating large many-particle systems whose potential energy is governed by two-particle contributions represents an important issue in computational and statistical physics.
A simple exemplary pair interaction model is given through the Lennard-Jones potential describing a novel gas using only inter-particle distances, whereas more complex {pair interactions} e.g. the Stockmayer potential~\cite{Stockmayer1941} or anisotropic Lennard-Jones potentials~\cite{Berne1972,Gay1981,Cleaver1996}
take the particles' orientation additionally into account.
Molecular or colloidal pair potentials may stem from various coarse-graining methods.
These methods are systematically treating microscopic details and dynamics on a coarser level of detail by integrating out “irrelevant” degrees of freedom.
{ 
For this purpose, different microscopic configurations of the system are grouped in terms of so called {\it coarse-grained states}.
A very simple and common example for a coarse-grained state is those grouping all microstates belonging to a specific center of mass distance between two molecules.
For such a specific state, an interaction can be assigned by means of a large series of sampled microstates using various coarse-graining methods.
%
%
Common methods in the literature include force matching schemes~\cite{Loewen1993,Ercolessi1994,Izvekov2005,Izvekov2005jcp},  the relative entropy method \cite{Shell2008}, the conditional reversible work method \cite{Brini2011},  the inverse Monte Carlo method\cite{Lyubartsev1995,Lyubartsev1997} or the iterative Boltzmann inversion method \cite{Soper1996,MullerPlathe2002} as well as hybrid schemes \cite{Ruhle2011}.
Besides these methods aiming to represent equilibrium quantities, there are efforts representing dynamical quantities as well.~\cite{Izvekov2006,Davtyan2015}
}
With regard to {all these} methods, there are many different fields of employment of which the following are just a few examples.
There are { simulation} studies investigating effective interactions, among alkanes \cite{Nielsen2003}, ring polymers \cite{Bernabei2013,Poier2014,Poier2015}, liquid hydrocarbons \cite{Jorgensen1984} or polycyclic aromatic hydrocarbons (PAH) \cite{Lilienfeld2006} and biomolecules \cite{Norberg1995,MorrissAndrews2009,MorrissAndrews2010,Izvekov2005,Bennun2009,Masunov2003,Villa2004,Bhattacherjee2016}.
Other challenging applications cover virus assembly simulations \cite{Hagan2016}
as well as simulations of colloidal systems\cite{Muller2014,Li2005}.

{Irrespective of the chosen method}, tables {characterizing} the coarse-grained states in terms of forces, energies or distributions are required.
{
From a computational perspective tables can easily lead to a memory storage problem since each new variable for the characterization of the coarse-grained state introduces a new dimension.
By assuming five variables - each having $100$ bins - we obtain an interaction table in a binary file format requiring around $37$ GB memory when using 32 bit data types.
The amount of data points comprising this table can thus easily become bigger than the number of sampled microstates (system snapshots) from the underlying simulation needed to calculate the effective pair potential.
Accordingly, very long simulation runs (in the milli- or micro second regime) are required for the creation of such an interaction table.
The first logical step to tackle this computational problem consists in avoiding redundant states in the table, i.e. we require each coarse-grained state to appear only once in the table. 
Tabulating interactions with the latter described optimal or non-redundant property also leads to a faster look-up time accordingly.

}

\begin{figure}[htb]
\begin{center}
 \includegraphics[width=8.5cm]{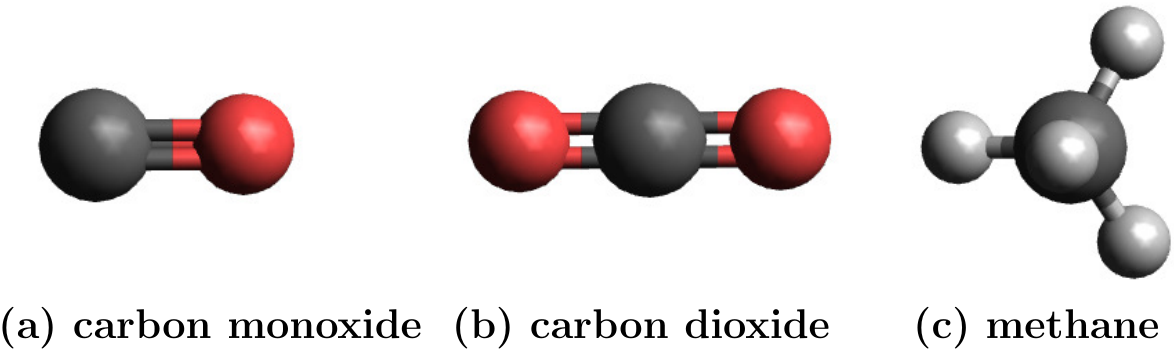}
\end{center}
 \caption{ Candidate molecules for (a) polar (b) uniaxial and (c) spherical anisotropy.
 Gray: carbon. Red: oxygen. White: hydrogen.}
 \label{fig:types0}
\end{figure}

{In our work, we investigate and provide appropriate sets of reaction coordinates for the description of coarse-grained states, fulfilling the non-redundant property, for pairs of mixed
molecular or colloidal particles displaying nearly spherical, polar (i.e., uniaxial without head-to-tail symmetry) 
or uniaxial kind of symmetry (Fig. \ref{fig:types0} shows an example for each). 
We hereby consider two- and three-dimensional systems.
}
By assuming these molecular symmetries and additional symmetries, we arrive at various definitions for coarse-grained states {in terms of appropriate reaction coordinates}.
The common case, where the inter-molecular distance is the only reaction coordinate, is used e.g. in star polymers\cite{Likos2001}.
However, extensions towards distance and orientation dependence have been conducted for protein segments \cite{Sippl1990,Buchete2004,Zhou2011}
and more recently for disc-shaped molecules \cite{Heinemann2014,Heinemannphd2016} and ring-shaped polymers \cite{Poier2015}. 
A further aspect in this work is to present a guideline for determining the effective inter-molecular pair potential out of atomistic detail.
We hereby follow the Boltzmann inversion principle and force the probability distribution as a function of the coarse-grained states { in equilibrium} to remain
invariant during the change from the microscopic to the coarse-grained level of detail.
{
This will lead us to an effective pair potential that is temperature dependent.
A very prominent method for gaining an effective pair potential is the iterative Boltzmann inversion scheme.
The latter is aiming to match the pair correlation function among coarse-grained sites of microscopic and coarse-grained detail, respectively.
In this work, however, we constrain to the special case that the entire system consists of only two molecules, each forming one coarse-grained site described through the molecular position and orientation.
This assumption is sufficient in systems of rather rigid molecules at low densities.\cite{Heinemann2014}
For this special case of a simple two-molecule system, an iterative procedure is no longer required and we can perform a direct Boltzmann inversion~\cite{Tschoep1998,Tschoep1998b}.
The corresponding potential is thus independent from the particle environment surrounding the two molecules.
}
%

Once having calculated tables representing effective interactions, simple models can be developed (e.g., as used for PAH molecules \cite{Heinemann2014,Heinemann2015,HernandezRojas2016}).
Despite the fact that these tables can become quite large, the direct use of {higher dimensional} numerical potentials in a sorted tabular form suitable for quick computational evaluations { was
already used for molecules\cite{Zacharopoulos2005,Lettieri2012,Spiriti2014,Spiriti2015}.}

At this point it is worthwhile to mention that describing the effective Hamiltonian using only single and pair contributions
represents a strong simplification in systems where inter-molecular interactions lead to strong polarization effects or conformational changes.
A discussion concerning this issue can be found in the work of A. A. Louis \cite{Louis2002}.
In his work he starts addressing the non-pairwise character of noble gas {possessing} a three-particle contribution from three-body Axilrod–Teller triple-dipole interactions\cite{Axilrod1943}. 
In general, the higher the density, the three particle contributions become more pronounced. Even though the three particle contribution for benzene in a crystalline phase is rather small, it is not neglectable~\cite{Evans1976}.
However, developing higher particle effective potentials requires a large number of rather complex reaction coordinates.
{As earlier mentioned}, the table describing the interactions is {becoming very large even if non-redundant reaction coordinates are chosen}.
Furthermore, the convergence of the partition sum in terms of effective many-particle Hamiltonians remains to be explored for the specific 
system being coarse-grained.
For these reasons, we here focus first on effective pair potentials.

\par
The remainder of this work is organized as follows. 
In Section \ref{sec:Defining the coarse-grained states in  a dimer} we define coarse-grained states formed by two molecules and provide in Sec. \ref{sec:Reaction coordinate description for coarse-grained states in a dimer} reaction coordinates whose sets of values correspond to these states via a biunique assignment.
We further {scrutinize} this biunique character in Section \ref{sec:Numerical evidence of the optimal choice of reaction coordinates}.
In order to obtain the interactions in form of numerical tables, Section \ref{sec:Computational methodology} provides a computational methodology {via} a step-by-step recipe.
This recipe is designed for the effective potential based on the Boltzmann inversion principle.
Finally, in Section \ref{sec:Conclusion} we summarize our findings.

\section{\label{sec:Defining the coarse-grained states in  a dimer}Defining the coarse-grained states in  a dimer}

The goal of the present section is to {provide definitions for} coarse-grained states for the dimer types that can be formed with our considered particle types (see Fig.~\ref{fig:types0}) in two and three dimensional systems.
{For this purpose, we provide a detailed investigation for coarse-grained states in two dimensional systems in Subsection \ref{sec:Coarse-grained states in two dimensions}} and extend this investigation to three dimensional systems in Subsection \ref{sec:Coarse-grained states in three dimensions}.
{
We later provide a description of the coarse-grained states in terms of reaction coordinates in Sec. \ref{sec:Reaction coordinate description for coarse-grained states in a dimer}. 
}

\subsection{\label{sec:Coarse-grained states in two dimensions}Coarse-grained states in two dimensions}

\begin{figure}[htb]
\begin{center}
 \includegraphics[width=5.5cm]{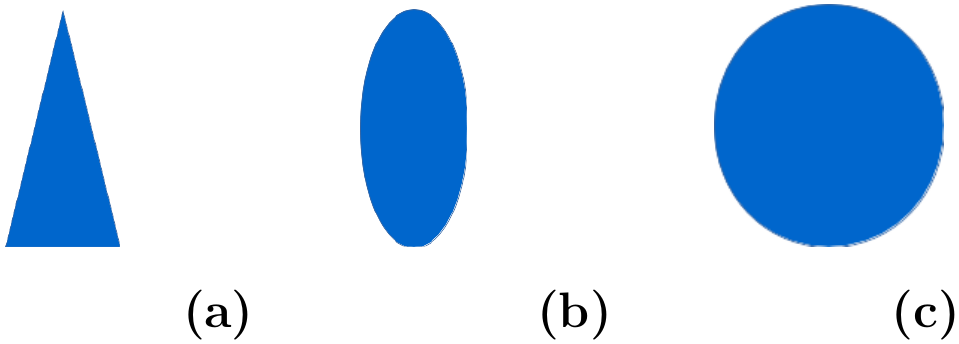}
\end{center}
 \caption{ Examples for dihedral particles (see main text) with the corresponding dihedral groups (a) $D_{1}$, (b) $D_2$ and (c) $D_{\infty}$.}
 \label{fig:types}
\end{figure}

\begin{figure}[htb]
\begin{center}
 \includegraphics[width=8.5cm]{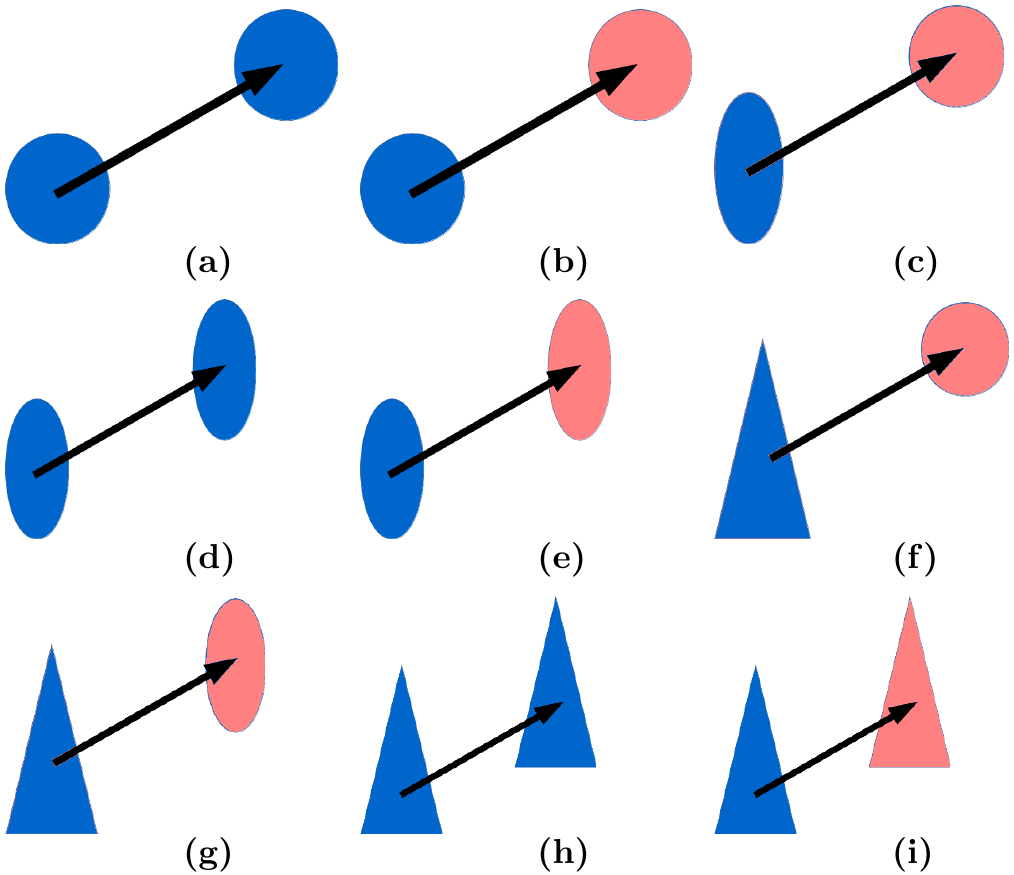}
\end{center} 
 \caption{ Various types of dimer configurations we further aim to describe in terms of optimal (non-redundant) reaction coordinates. These configurations differ in the particle's (molecule's) dihedral group and species. (a) $D_{\infty}-D_{\infty}$, (b) $D_{\infty}-\underline{D}_{\infty}$, (c) $D_2-D_{\infty}$, (d) $D_2-D_2$, (e) $D_2-\underline{D}_2$, (f) $D_1-D_{\infty}$, (g) $D_1-D_2$, (h) $D_1-D_1$, (i) $D_1-\underline{D}_1$. A different color (or an underlined letter) denotes that particles (molecules) are {distinct}.}
 \label{fig:rcpics}
\end{figure}

To proceed, we next introduce a simple notation for the particle types in two dimensions.
In particular, polar, uniaxial and circular particles correspond to dihedral ($D$) particles of first, second and infinite order (indicated by subscript).
A dihedral particle of first order ($D_1$) owns one reflective axis and has no rotational symmetry (rotation group $R_1$).
Further, a dihedral particle of second order ($D_2$) owns two reflective axes and its pattern can be reproduced by rotating the particle about $180$ degrees ($R_2$).
A circular particle ($D_{\infty}$) marks the limit of infinite reflective axes.
Examples are sketched in Fig.~\ref{fig:types}.
{In our investigation, we focus} on dimer systems with mixed particle symmetries (see Fig. \ref{fig:rcpics}).

\begin{figure}[htb]
\begin{center}
 \includegraphics[width=4.5cm]{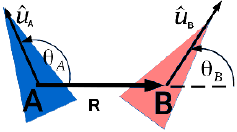}
\end{center}
 \caption{ Schematic view of a two-dimensional dimer comprising particle (or molecule) A and B with orientation vector $\hat{\mathbf{u}}_{\rm A}$ and  $\hat{\mathbf{u}}_{\rm B}$.
 The angles $\theta_{\rm A}$ and $\theta_{\rm B}$ describe the tilt angle of the particles with respect to their inter-particle connecting vector $\mathbf{R}$.
}
 \label{fig:dimer}
\end{figure}

We next parametrize each of the two molecules A(B) of a dimer via a position vector $\mathbf{R}_{\rm A(B)}$ and one normalized orientation vector $\hat{\mathbf{u}}_{\rm A(B)}$.
Then, we extract three scalar parameters out of this set of four vectors and introduce symmetry operations in these scalar variables.
These operations classify the type of the particle dimer.
In particular, we define the inter-molecular distance $R$ and the variables $\theta_{\rm A}$ and $\theta_{\rm B}$ denoting the angles (in positive direction) from the inter-molecular
distance vector (pointing from molecule A to B) and the corresponding orientation vector of molecule A and B, that is,
\begin{subequations}
\allowdisplaybreaks
\label{eqn:rcac}
\begin{align}
R&=\left|\mathbf{R}\right|=\left|\mathbf{R}_{\text{B}}-\mathbf{R}_{\text{A}}\right| \label{eqn:rca}\\
\theta_{\rm A}&=\measuredangle (\mathbf{R}, \mathbf{\hat{u}}_{\text{A}})
\label{eqn:rcb}
\\
\theta_{\rm B}&=\measuredangle (\mathbf{R}, \mathbf{\hat{u}}_{\text{B}})\text{.}
\label{eqn:rcc}
\end{align}
\end{subequations}
A graphical representation of the latter described parametrization is depicted in Fig. \ref{fig:dimer}.
In the following, we define functions that characterize the type of the molecular dimer with respect to the symmetries which the molecules possess by approximation (see second column in Table \ref{tab:2d3d} on page \pageref{tab:2d3d}).
\begin{subequations}
\allowdisplaybreaks
\label{eqn:symmetries2d}
\begin{align}
\text{Circular symmetry:}\qquad\qquad&\nonumber\\
 S_{\rm A}^x: (R, \theta_{\rm A},\theta_{\rm B})&\mapsto (R, \theta_{\rm A}+x,\theta_{\rm B}), \qquad \forall x\\
 S_{\rm B}^x: (R, \theta_{\rm A},\theta_{\rm B})&\mapsto (R, \theta_{\rm A},\theta_{\rm B}+x), \qquad \forall x\\
\text{\centering{Head-to-tail symmetry:}}\qquad\qquad&\nonumber\\
%
 F_{\rm A}: (R, \theta_{\rm A},\theta_{\rm B})&\mapsto (R, \theta_{\rm A}+\pi,\theta_{\rm B})\\
 F_{\rm B}: (R, \theta_{\rm A},\theta_{\rm B})&\mapsto (R, \theta_{\rm A},\theta_{\rm B}+\pi)\\
%
\text{\centering{Chiral symmetry (invariant when mirroring): }}&\nonumber\\
%
\chi: (R, \theta_{\rm A},\theta_{\rm B})&\mapsto (R, -\theta_{\rm A},-\theta_{\rm B})\\
%
\text{\centering{Particles are indistinguishable:}}&\nonumber\\
%
V: (R, \theta_{\rm A},\theta_{\rm B})&\mapsto (R, \pi-\theta_{\rm B},\pi-\theta_{\rm A})\\
%
{\text{\centering{Angular periodicity:}}}&\nonumber\\
%
{\alpha: (R, \theta_{\rm A},\theta_{\rm B})}&{\mapsto (R, \text{Arg}(e^{\imath\theta_{\rm A}}),\text{Arg}(e^{\imath\theta_{\rm B}}))}
\end{align}
\end{subequations}
It is further possible to group parametrizations such as $R',\theta_{\rm A}',\theta_{\rm B}'$ and $R'',\theta_{\rm A}'',\theta_{\rm B}''$, that form indistinguishable configurations with respect to their symmetry operations, into one equivalence class denoted by $[R',\theta_{\rm A}',\theta_{\rm B}']$ or $[R'',\theta_{\rm A}'',\theta_{\rm B}'']$.
{
For this purpose we introduce the new set of symmetry functions according to the following scheme:
\begin{align}
\label{eqn:tilde}
\tilde{S}=\alpha \circ{} S \circ{} \alpha, S\in \{F_{\rm A}, F_{\rm B}, \chi, V, S_{\rm A}^{x \in \mathbb{R}}, S_{\rm B}^{x \in \mathbb{R}} \} 
\end{align}
We define an equivalence class that forms a coarse-grained state through
\begin{multline}
\label{eqn:2dcgstate}
[R, \theta_{\rm A},\theta_{\rm B}]=\big\{(R',\theta_{\rm A}',\theta_{\rm B}') \big| \alpha (R,\theta_{\rm A},\theta_{\rm B}) = \\ \tilde{S}_1 \circ \dots \circ \tilde{S}_N (R',\theta_{\rm A}',\theta_{\rm B}'), N\in \mathbb{N}, \tilde{S}_i \in \{ \tilde{F}_{\rm A},\tilde{F}_{\rm B},\tilde{\chi},\tilde{V}, \tilde{S}_{\rm A}^{x \in \mathbb{R}}, \tilde{S}_{\rm B}^{x \in \mathbb{R}} \} \big\}. 
\end{multline}
}

{
This definition suggests that a finite but generally large number  $N\in \mathbb{N}$ of operations is needed to map one set of parameters $(R',\theta_{\rm A}',\theta_{\rm B}')$
onto another set of parameters $\alpha(R,\theta_{\rm A},\theta_{\rm B})$ in the same equivalence class.
This number can be significantly shortened using the following relations.

\begin{subequations}
\begin{align}
\allowdisplaybreaks
\label{eqn:relationsa}
\tilde{F}_{\rm A}\circ{} \tilde{F}_{\rm A}=\tilde{F}_{\rm B}\circ{} \tilde{F}_{\rm B}&=\tilde{\chi}\circ{}\tilde{\chi}=\tilde{V}\circ{}\tilde{V}=\alpha\\ 
\tilde{V} \circ{} \tilde{F}_{\rm A/B}&=\tilde{F}_{\rm B/A} \circ{} \tilde{V}\\ 
[\tilde{F}_{\rm A},\tilde{F}_{\rm B}]=[\tilde{F}_{\rm A},\tilde{\chi}]&=[\tilde{F}_{\rm B},\tilde{\chi}]=[\tilde{\chi},\tilde{V}]=0
\label{eqn:relationsc}
\end{align}
\end{subequations}

}

\subsection{\label{sec:Coarse-grained states in three dimensions}Coarse-grained states in three dimensions}

\begin{figure}[htb]
\begin{center}
 \includegraphics[width=4cm]{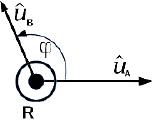}
\end{center}
 \caption{Schematic view of a dimer in three dimensions along the inter-particle (inter-molecular) connecting vector $\mathbf{R}$. The angle $\phi$ describes the right-handed twist of the orientation vector $\hat{\mathbf{u}}_{\rm B}$ of particle (or molecule) B with respect to the orientation vector $\hat{\mathbf{u}}_{\rm A}$ of particle A along the vector $\mathbf{R}$.}
 \label{fig:dimer3d}
\end{figure}

In analogy to the two-dimensional dimer systems, we can introduce three-dimensional dimer systems where the particles are axially symmetric around their orientation vector.
Specifically, we adopt the parametrization from the Eqs. \eqref{eqn:rca}--\eqref{eqn:rcc}
and additionally consider the angle 
\begin{align}
\phi=\measuredangle (\mathbf{\hat{u}}_{\text{A}}-\mathbf{\hat{R}}(\mathbf{\hat{u}}_{\text{A}}\cdot \mathbf{\hat{R}}),\mathbf{\hat{u}}_{\text{B}}-\mathbf{\hat{R}}(\mathbf{\hat{u}}_{\text{B}}\cdot \mathbf{\hat{R}}))\text{,}
\label{eqn:rcphi}
\end{align}
which denotes the right-handed torsion angle between the projections of the orientation vectors perpendicular to $\mathbf{R}$ (or its normalized counterpart $\mathbf{\hat{R}}$) as depicted in Fig. \ref{fig:dimer3d}.
The symmetries from previous paragraph are characterized through the following functions:
\begin{subequations}
\allowdisplaybreaks
\label{eqn:symmetries3d}
\begin{align}
\text{Spherical symmetry:}\qquad\qquad&\nonumber\\
 S_{\rm A}^{x,y}: (R, \theta_{\rm A},\theta_{\rm B},\phi)&\mapsto (R, \theta_{\rm A}+x,\theta_{\rm B},\phi+y), \qquad \forall x,y\\
 S_{\rm B}^{x,y}: (R, \theta_{\rm A},\theta_{\rm B},\phi)&\mapsto (R, \theta_{\rm A},\theta_{\rm B}+x,\phi+y), \qquad \forall x,y\\
%
\text{Head-to-tail symmetry:}\qquad\qquad&\nonumber\\
%
 F_{\rm A}: \left(R, \theta_{\rm A},\theta_{\rm B},\phi \right)&\mapsto \left(R,  \pi-\theta_{\rm A},\theta_{\rm B},\phi+\pi\right)\\
 F_{\rm B}: \left(R, \theta_{\rm A},\theta_{\rm B},\phi \right)&\mapsto \left(R,  \theta_{\rm A},\pi-\theta_{\rm B},\phi+\pi\right)\\
%
\text{Chiral symmetry:}\qquad\qquad&\nonumber\\
 \chi:\left(R, \theta_{\rm A},\theta_{\rm B},\phi \right)&\mapsto \left(R,  \theta_{\rm A},\theta_{\rm B},-\phi\right)\\
%
\text{Particles are indistinguishable:}\qquad\qquad&\nonumber\\
%
V: \left(R, \theta_{\rm A},\theta_{\rm B},\phi \right)&\mapsto \left(R,  \pi-\theta_{\rm B},\pi-\theta_{\rm A},-\phi\right)\\
%
{\text{\centering{Angular periodicity:}}}\qquad\qquad&\nonumber\\
%
{\alpha: (R, \theta_{\rm A},\theta_{\rm B},\phi)}&{\mapsto \left(R, \arccos \cos(\theta_{\rm A}), \arccos \cos(\theta_{\rm B}), 
\text{Arg}(e^{\imath\cdot {\Phi} }    )
\right),}
\\
&{\text{with} \; \Phi=\begin{cases}\pi-\phi, \;\text{sgn}(\sin(\theta_{\rm A})\cdot \sin(\theta_{\rm B}))<0
 \\\phi ,\; \text{else}\end{cases}  }   \nonumber
\end{align}
\end{subequations}
{
The coarse-grained states in three dimensional systems are defined in analogy to Eq. \eqref{eqn:2dcgstate} through the following expression
\begin{multline}
\label{eqn:3dcgstate}
[R, \theta_{\rm A},\theta_{\rm B},\phi]=\big\{(R',\theta_{\rm A}',\theta_{\rm B}',\phi') \big| \alpha(R,\theta_{\rm A},\theta_{\rm B},\phi) = \\ \tilde{S}_1 \circ \dots \circ \tilde{S}_N (R',\theta_{\rm A}',\theta_{\rm B}',\phi'), N\in \mathbb{N}, \tilde{S}_i \in \{ \tilde{F}_{\rm A},\tilde{F}_{\rm B},\chi,\tilde{V}, \tilde{S}_{\rm A}^{x,y  \in \mathbb{R}}, \tilde{S}_{\rm B}^{x,y  \in \mathbb{R}} \} \big\}. 
\end{multline}
}

\section{\label{sec:Reaction coordinate description for coarse-grained states in a dimer}Reaction coordinate description for coarse-grained states in a dimer}
{
In this section we present reaction coordinates that describe a one-to-one correspondence between the set of equivalence classes (representing the coarse-grained states) and the value set of these reaction coordinates, which we denote with $\mathfrak{Q}$.
The latter is thus minimal in size.
Tabulated functions in these optimal or non-redundant reaction coordinates require a reduced memory space and look-up time accordingly. 
Moreover, the amount of simulation data required to produce these tabulated functions also reduces. This reduction is obviously proportional to the table size reduction itself.
The one-to-one correspondence can be expressed via a bijective function $f$ for each dimer type, fulfilling
\begin{align}\label{eqn:f1}
 \text{CG\,states} \;\xrightarrow{\text{bijective\, function}\, f} \;\text{reaction \,coordinate space $\mathfrak{Q}$}.
\end{align}
with $f$ being
\begin{align}\label{eqn:f2}
f : [R,\theta_{\rm A},\theta_{\rm B} (,\phi)] \mapsto \left( \begin{array}{c} \bar{q}_1(R, \theta_{\rm A},\theta_{\rm B} (,\phi))\\\bar{q}_2(R,\theta_{\rm A},\theta_{\rm B} (,\phi)) \\ \vdots \\ \bar{q}_M(R,\theta_{\rm A},\theta_{\rm B} (,\phi)) \end{array} \right)\text{,}
\end{align}
where the $\bar{q}_i, i=1,\dots,M$ denote the mapping functions that map each parametrization $R, \theta_{\rm A},\theta_{\rm B} (,\phi)$ onto the corresponding reaction coordinate value.
This function $f$ is bijective when being injective, i.e.
\begin{align}
f(R,\theta_{\rm A},\theta_{\rm B} (, \phi)) = f(R',\theta_{\rm A}',\theta_{\rm B}' (, \phi')) \;  \Rightarrow \;  [R,\theta_{\rm A},\theta_{\rm B} (, \phi)]=[R',\theta_{\rm A}',\theta_{\rm B}' (, \phi')]
\label{eqn:injective}
\end{align}
and surjective, i.e.
\begin{align}
\forall (q_1,\dots,q_M)^{T} \in \mathfrak{Q} \, \exists \, R, \theta_{\rm A}, \theta_{\rm B} (, \phi) \, \textrm{with} \,f(R, \theta_{\rm A}, \theta_{\rm B} (, \phi))=(q_1,\dots,q_M)^{T}.
\label{eqn:surjective}
\end{align}
In order to permit calculations of distribution functions in our later proposed reaction coordinates, we have to force that small deviations in the parametrization $R, \theta_{\rm A},\theta_{\rm B} (,\phi)$ lead to 
small deviations in the mapped reaction coordinate values.
In other words, the mapping functions $\bar{q}_i$  have to be continuous in their arguments.
Fortunately, we found sets of $\bar{q}_i$ that fulfill this criterion (see third and fourth column of Table \ref{tab:2d3d} on page \pageref{tab:2d3d}) except for
one mapping function ($\bar{q}_4$ in the third column) exhibiting jumps if the argument in the $\text{sgn}_{+}$ function (defined in the caption of Tab. \ref{tab:2d3d}) switches sign.
Since these jumps form only a zero set in the space formed by the parametrization variables ($R, \theta_{\rm A}, \theta_{\rm B}$) the occurrence is thus rather unlikely.

It is further worth mentioning that the bijective character of $f$ for all dimer types can be proven. We, however, provide one exemplary proof for the $D_1-D_2$ dimer in two dimension in Appendix \ref{sec:Exemplary proof of f being bijective}.
Rather than providing the remaining proofs which are performed in a similar fashion, we check in Sec. \ref{sec:Numerical evidence of the optimal choice of reaction coordinates} the bijective character of $f$ numerically for all non-trivial dimer types and thereby analyze the sampling of the reaction coordinate space.

As we have pointed out, the most significant outcome of this choice of reaction coordinates is the minimal value set.
Furthermore, we quantify in Table \ref{tab:2d3defficiency} the value set reduction when going from the parametrization to the reaction coordinate description.
We detect for the $D_2-D_2$ dimer the highest value set reduction with an effective reduction of $\mathfrak{Q}$ down to a 16th in size.
In addition, we have quantified the value set reduction when using the reaction coordinates proposed by Berne and Pechukas for polar particles~\cite{Berne1972} in three dimensions.
The coordinates from Berne and Pechukas, however, only lead to a reduction of one-half since these only account for chiral symmetry.
We also want to point out that in our sets of reaction coordinates, the values for each reaction coordinate do not or only trivially constrain the allowed value sets of the other reaction coordinates.

}

\begin{table}[htb]
\caption{\label{tab:2d3d}
Collection of mapping functions describing a mapping between molecular angles (see Fig. \ref{fig:dimer} and Fig. \ref{fig:dimer3d}) and reaction coordinates for different molecular dimer types (Fig. \ref{fig:rcpics}).
Symmetries as described in Sec. \ref{sec:Defining the coarse-grained states in  a dimer} are implicitly included.
The following functions are used: $W(t)=\arccos(\cos t)/\pi$, $w(t,u,v)=\arccos(\cos(t)\  \cdot \text{sgn}(\cos(u)\cdot \cos(v)))/\pi$, $\mathfrak{W}(t)=2/\pi\cdot\arccos|\cos t|$, $\text{sgn}_+(t)=\text{sgn}(\text{sgn}(t)+1)$.
{Function $J$ (see Eq. \eqref{eqn:J}) characterizes the configuration space density as a function of the reaction coordinates.}
It is important to note, that the different powers of $R$ in the $\bar{q}_1$-functions do not influence the bijective property of $f$ (later introduced in Sec. \ref{sec:Numerical evidence of the optimal choice of reaction coordinates}), but lead to an equal sampling distribution along $q_1$ [for $\bar{q}_1=R \Rightarrow J\propto R^2$].
}
{\scalebox{0.67}{
\setstretch{0.9}
\begin{tabular}{ccll}
&&&\vspace*{-0.9em}\\
dimer type&symmetries&- 2-dim. mapping functions $\bar{q}_i(R,\theta_{\rm A},\theta_{\rm B})$ &- 3-dim. mapping functions $\bar{q}_i(R, \theta_{\rm A},\theta_{\rm B},\phi)$\\
see Fig. \ref{fig:rcpics})&&{\;\;with}&{\;\;with}\\
&&{\;\;$(R,\theta_{\rm A},\theta_{\rm B})\in ([0,\infty), [-\pi,\pi], [-\pi,\pi])$}&{\;\;$(R,\theta_{\rm A},\theta_{\rm B},\phi)\in ([0,\infty), [0,\pi], [0,\pi], [-\pi,\pi])$}\\
&&- $J(q_1,\dots,q_4)$&- $J(q_1,\dots,q_4)$\\
\hline
&&&\vspace*{-0.9em}\\
(a) $D_{\infty}-D_{\infty}$&$\tilde{F}_{\rm A},\tilde{F}_{\rm B}, \tilde{\chi},\tilde{V}$&$\bar{q}_1=R^2\in (0,\infty)$&$\bar{q}_1=R^3\in (0,\infty)$\\
&$\tilde{S}_{\rm A}, \tilde{S}_{\rm B}$&$J=const$&$J=const$\\
\hline
&&&\vspace*{-0.9em}\\
(b) $D_{\infty}-\underline{D}_{\infty}$&$\tilde{F}_{\rm A},\tilde{F}_{\rm B},\tilde{\chi},\tilde{V}$&$\bar{q}_1=R^2$&$\bar{q}_1=R^3$\\
&$\tilde{S}_{\rm A}, \tilde{S}_{\rm B}$&$J=const$&$J=const$\\
\hline
&&&\vspace*{-0.9em}\\
(c) $D_2-D_{\infty}$ & $\tilde{F}_{\rm A},\tilde{F}_{\rm B},\tilde{\chi}, \tilde{S}_{\rm B}$&$\bar{q}_1=R^2$&$\bar{q}_1=R^3$\\
&&$\bar{q}_2=\mathfrak{W}(\theta_{\rm A}) \in [0,1]$ & $\bar{q}_2=\left|\cos\left(\theta_{\rm A} \right)\right| \in [0,1]$\\
&&$J=const$&$J=const$\\
\hline
&&&\vspace*{-0.9em}\\
(d) $D_2-D_2$ & $\tilde{F}_{\rm A},\tilde{F}_{\rm B},\tilde{\chi},\tilde{V}$ &$\bar{q}_1=R^2$ &$\bar{q}_1=R^3$\\
&&$\bar{q}_2=\text{min}\left(\mathfrak{W}(\theta_{\rm A}),\mathfrak{W}(\theta_{\rm B})\right) \in [0,\bar{q}_3]$&$\bar{q}_2=\text{min}\left( |\cos(\theta_{\rm A})|,|\cos(\theta_{\rm B})|  \right) \in [0,\bar{q}_3]$\\
&&$\bar{q}_3=\text{max}\left( \mathfrak{W}(\theta_{\rm A}),\mathfrak{W}(\theta_{\rm B})\right) \in [\bar{q}_2,1]$&$\bar{q}_3=\text{max}\left( |\cos(\theta_{\rm A})|,|\cos(\theta_{\rm B})|  \right) \in [\bar{q}_2,1]$\\
&&$\bar{q}_4=\text{sgn}_{+}(\sin(2\theta_{\rm A})\cdot\sin(2\theta_{\rm B})) \in \{0,1\}$&$\bar{q}_4=w(\phi,\theta_{\rm A},\theta_{\rm B})\in [0,1]$\\ 
&&$J\propto\int_0^{\pi} \mathrm{d}\theta_{\rm A}\,\int_0^{\pi} \mathrm{d}\theta_{\rm B}  \,\delta(q_2-\bar{q}_2(\theta_{\rm A},\theta_{\rm B}))$&$J\propto\int_0^{\pi} \mathrm{d}\theta_{\rm A}\,\sin \theta_{\rm A}\,\int_0^{\pi} \mathrm{d}\theta_{\rm B}\, \sin\theta_{\rm B}  $\\
&&$\;\;\;\;\;\;\;\;\;\;\;\;\;\;\;\;\;\;\;\;\;\;\;\;\;\;\;\;\cdot \delta(q_3-\bar{q}_3(\theta_{\rm A},\theta_{\rm B}))$&$\;\;\;\;\;\;\;\;\cdot \delta(q_2-\bar{q}_2(\theta_{\rm A},\theta_{\rm B}))\, \delta(q_3-\bar{q}_3(\theta_{\rm A},\theta_{\rm B}))$\\
\hline
&&&\vspace*{-0.9em}\\
(e) $D_2-\underline{D}_2$ & $\tilde{F}_{\rm A},\tilde{F}_{\rm B},\tilde{\chi}$ & $\bar{q}_1=R^2$ & $\bar{q}_1=R^3$\\
&&$\bar{q}_2=\mathfrak{W}(\theta_{\rm A}) \in [0,1]$&$\bar{q}_2=|\cos(\theta_{\rm A})| \in [0,1]$\\
&&$\bar{q}_3=\mathfrak{W}(\theta_{\rm B}) \in [0,1]$&$\bar{q}_3=|\cos(\theta_{\rm B})| \in [0,1]$\\
&&$\bar{q}_4=\text{sgn}_{+}(\sin(2\theta_{\rm A})\cdot\sin(2\theta_{\rm B})) \in\{0,1\}$&$\bar{q}_4=w(\phi,\theta_{\rm A},\theta_{\rm B})\in [0,1]$\\ 
&&$J=const$&$J=const$\\
\hline
&&&\vspace*{-0.9em}\\
(f) $D_1-D_{\infty}$ & $\tilde{F}_{\rm B},\tilde{\chi}, \tilde{S}_{\rm B}$&$\bar{q}_1=R^2$&$\bar{q}_1=R^3$\\
&&$\bar{q}_2=W(\theta_{\rm A})\in [0,1]$&$\bar{q}_2=\cos\left(\theta_{\rm A} \right)\in [-1,1]$\\
&&$J=const$&$J=const$\\
\hline
&&&\vspace*{-0.9em}\\
(g) $D_1-D_2$ & $\tilde{F}_{\rm B},\tilde{\chi}$ & $\bar{q}_1=R^2$ & $\bar{q}_1=R^3$\\
&&$\bar{q}_2=W(\theta_{\rm A})\in [0,1]$&$\bar{q}_2=\cos(\theta_{\rm A})\in [-1,1]$\\
&&$\bar{q}_3=\mathfrak{W}(\theta_{\rm B})\in [0,1]$&$\bar{q}_3=|\cos(\theta_{\rm B})|\in [0,1]$\\
&&$\bar{q}_4=\text{sgn}_{+}(\sin(\theta_{\rm A})\cdot\sin(2\theta_{\rm B}))\in \{0,1\}$&$\bar{q}_4=w(\phi,0,\theta_{\rm B})\in [0,1]$\\ 
&&$J=const$&$J=const$\\
\hline
&&&\vspace*{-0.9em}\\
(h) $D_1-D_1$ & $\tilde{\chi}, \tilde{V}$&$\bar{q}_1=R^2$&$\bar{q}_1=R^3$\\
&&$\bar{q}_2=\text{min}\left( W(\theta_{\rm A}),W(\pi-\theta_{\rm B})  \right)\in [0,\bar{q}_3]$& $\bar{q}_2=\text{min}\left( \cos(\theta_{\rm A}),-\cos(\theta_{\rm B})  \right)\in [-1,\bar{q}_3]$\\
&&$\bar{q}_3=\text{max}\left( W(\theta_{\rm A}),W(\pi-\theta_{\rm B})  \right)\in [\bar{q}_2,1]$&$\bar{q}_3=\text{max}\left( \cos(\theta_{\rm A}),-\cos(\theta_{\rm B})  \right)\in [\bar{q}_2,1]$\\
&&$\bar{q}_4=\text{sgn}_{+}(\sin(\theta_{\rm A})\cdot\sin(\theta_{\rm B})) \in \{0,1\}$&$\bar{q}_4=W(\phi) \in [0,1]$\\ 
&&$J\propto\int_0^{\pi} \mathrm{d}\theta_{\rm A}\,\int_0^{\pi} \mathrm{d}\theta_{\rm B}  \,\delta(q_2-\bar{q}_2(\theta_{\rm A},\theta_{\rm B}))$&$J\propto\int_0^{\pi} \mathrm{d}\theta_{\rm A}\,\sin \theta_{\rm A}\,\int_0^{\pi} \mathrm{d}\theta_{\rm B}\, \sin\theta_{\rm B}  $\\
&&$\;\;\;\;\;\;\;\;\;\;\;\;\;\;\;\;\;\;\;\;\;\;\;\;\;\;\;\;\cdot \delta(q_3-\bar{q}_3(\theta_{\rm A},\theta_{\rm B}))$&$\;\;\;\;\;\;\;\;\cdot \delta(q_2-\bar{q}_2(\theta_{\rm A},\theta_{\rm B}))\, \delta(q_3-\bar{q}_3(\theta_{\rm A},\theta_{\rm B}))$\\
\hline
&&&\vspace*{-0.9em}\\
(i) $D_1-\underline{D}_1$ & $\tilde{\chi}$ &$\bar{q}_1=R^2$ &$\bar{q}_1=R^3$\\
&&$\bar{q}_2= W(\theta_{\rm A}) \in [0,1]$&$\bar{q}_2= \cos(\theta_{\rm A}) \in [-1,1]$\\
&&$\bar{q}_3=W(\theta_{\rm B}) \in [0,1]$&$\bar{q}_3=\cos(\theta_{\rm B}) \in [-1,1]$\\
&&$\bar{q}_4=\text{sgn}_{+}(\sin(\theta_{\rm A})\cdot\sin(\theta_{\rm B})) \in \{0,1\}$&$\bar{q}_4=W(\phi) \in [0,1]$\\
&&$J=const$&$J=const$
\end{tabular}
}}
\end{table}

\begin{table}[htb]
\caption{\label{tab:2d3defficiency}
Value set reduction factor for the parametrization given through Eqs. \eqref{eqn:rcac} (for three dimensions also Eq. \eqref{eqn:rcphi}) shown when using the symmetries [depending on the dimer type  (see Fig. \ref{fig:rcpics})] from Eqs. \eqref{eqn:symmetries2d} (for three dimensions: Eqs. \eqref{eqn:symmetries3d}).
We hereby show results for our reaction coordinates (rc's) (see Table \ref{tab:2d3d}) and those used by Berne and Pechukas \cite{Berne1972}.
Each of the additional symmetries ($\tilde{F}_{\rm A},\tilde{F}_{\rm B},\tilde{\chi}$ or $\tilde{V}$) reduces the value set by a half, {since each symmetry relation identifies two parametrizations $(R,\theta_{\rm A},\theta_{\rm B} (,\phi)) \equiv (R',\theta_{\rm A}',\theta_{\rm B}' (,\phi'))$ which cannot be identified using any combination of the other symmetry relations.}
}
\scalebox{0.8}{
\setstretch {1}
\begin{tabular}{ccc}
&&\vspace*{-0.9em}\\
dimer type& additional symmetries  & additional symmetries\\
(see Fig. \ref{fig:rcpics})& using our rc's& using Berne-Pechukas rc's \\
&(value set scaling factor&(value set scaling factor\\
& for 2 and 3 dimensions)& for 3 dimensions)\\
\hline
&&\vspace*{-0.9em}\\
(d) $D_2-D_2$ &$\tilde{F}_{\rm A},\tilde{F}_{\rm B},\tilde{\chi},\tilde{V}$ \,$(1/16)$&$\tilde{\chi}$ \,$(1/2)$ \\
\hline
&&\vspace*{-0.9em}\\
(e) $D_2-\underline{D}_2$ &$\tilde{F}_{\rm A},\tilde{F}_{\rm B},\tilde{\chi}$ \,$(1/8)$&$\tilde{\chi}$ \,$(1/2)$\\
\hline
&&\vspace*{-0.9em}\\
(g) $D_1-D_2$ &$\tilde{F}_{\rm B},\tilde{\chi}$ \,$(1/4)$&$\tilde{\chi}$ \,$(1/2)$\\
\hline
&&\vspace*{-0.9em}\\
(h) $D_1-D_1$ &$\tilde{\chi},\tilde{V}$ \,$(1/4)$&$\tilde{\chi}$ \,$(1/2)$\\
\hline
&&\vspace*{-0.9em}\\
(i) $D_1-\underline{D}_1$ &$\tilde{\chi}$ \,$(1/2)$&$\tilde{\chi}$ \,$(1/2)$\\
\end{tabular}
}
\end{table}

{
\section{\label{sec:Numerical evidence of the optimal choice of reaction coordinates}   Numerical evidence of the optimal choice of reaction coordinates  }

We next provide numerical evidence of $f$ being bijective for  the non-trivial dimer symmetries given by the subfigures (d), (e), (g), (h) and (i) of Fig. \ref{fig:rcpics}. 
For this purpose, we introduce the following discretizations that reflect an evenly sampling of the orientation space of both particles with vectors $\mathbf{\hat{u}}_{\rm A}$ and $\mathbf{\hat{u}}_{\rm B}$ in the body-fixed dimer systems illustrated through Fig. \ref{fig:dimer} and Fig. \ref{fig:dimer3d} for two-dimensional systems:
\begin{subequations}
 \allowdisplaybreaks
  \begin{align}
(\theta_{\rm A})_i &=2\pi/1440\cdot i -\pi \qquad , i\in\{0,1439\}  
\label{eqn:samp2da}
\\
(\theta_{\rm B})_i &=2\pi/1440\cdot i -\pi \qquad , i\in\{0,1439\}  
\label{eqn:samp2db}
\end{align}
\end{subequations}
and for three-dimensional systems:
\begin{subequations}
 \allowdisplaybreaks
 \begin{align}
(\cos(\theta_{\rm A}))_i &=2/540 \cdot i-1 \qquad , i\in\{0,539\}
\label{eqn:samp3da}
\\
(\cos(\theta_{\rm B}))_i &=2/540 \cdot i-1 \qquad , i\in\{0,539\}\\
\phi_i&=2\pi/720\cdot i -\pi \qquad , i\in\{0,719\}
\label{eqn:samp3dc}
\end{align}
\end{subequations}
For the these sets of points we calculated values of the reaction coordinates (see Table \ref{tab:2d3d}) for the different dimer symmetries and stored them together with the underlying angles in a large file.
The rows of this file have the following form for the two-dimensional case:
$$\theta_{\rm A}, \theta_{\rm B}, \bar{q}_2(\theta_{\rm A}, \theta_{\rm B}), \bar{q}_3(\theta_{\rm A}, \theta_{\rm B}), \bar{q}_4(\theta_{\rm A}, \theta_{\rm B})$$
and the three-dimensional case:
$$\theta_{\rm A}, \theta_{\rm B}, \phi, \bar{q}_2(\theta_{\rm A}, \theta_{\rm B}, \phi), \bar{q}_3(\theta_{\rm A}, \theta_{\rm B}, \phi), \bar{q}_4(\theta_{\rm A}, \theta_{\rm B}, \phi)$$
We sort this file with respect to the reaction coordinate values to get blocks of angles yielding the same  reaction coordinate values.
We checked each set of angles in a block whether it can be transformed into the first set of angles in this block using the symmetry functions $\tilde{F}_{\rm A}$, $\tilde{F}_{\rm B}$, $\tilde{\chi}$ and $\tilde{V}$ defined in Eqs. \eqref{eqn:symmetries2d}  or \eqref{eqn:symmetries3d} (in conjunction with Eq. \eqref{eqn:tilde}).

We can conclude from our calculations that each set of the sampled reaction coordinate values is created by angular configurations belonging to one distinct equivalence class.
Accordingly, the assignment between our detected set of equivalence classes and the set of all values of the reaction coordinates is injective (see Def. \eqref{eqn:injective}).


To numerically check the surjective character of $f$ (see Def. \eqref{eqn:surjective}) we first divided each reaction coordinate into many equal bins and then denoted the sampling rate of the combined bins of all orientation dependent reaction coordinates ($q_2$ to $q_4$).
{
Since the underlying sampling evenly samples the orientation space of the two orientation vectors in the dimer, the sampling rate of each combined bin has to be proportional to
the configuration space {density}.
}
In Figs. \ref{fig:rcsamplingdist2d} and \ref{fig:rcsamplingdist3d} we present the sampling rate as a function of the sorted bin number (for the two- and three-dimensional case).
Hereby, the bins are sorted by the sampling rate in ascending order. A dotted black line marks the highest sampled bin number.
We can conclude that in all cases all bins were sampled. 

Another issue worthwhile to mention is the sampling rate of each bin.
It is often desired to have reaction coordinates, where an equidistant binning or tabulating leads to the same sampling rate if no interaction among the particles is present.
Obviously this equal sampling is a consequence of a constant configuration space density which we denote with $J$ and explicitly define later in Sec. \ref{sec:Definition of the effective pair potential}.
As expected, we observe equal sampling for all cases where $J$ is constant. 
However, considering the cases (d) and (h), where both particles are indistinguishable, the 
sampling rate for our set of reaction coordinates is not equal for all bins as expected.

\begin{figure}[htb]
\begin{center}
 \includegraphics[width=12cm]{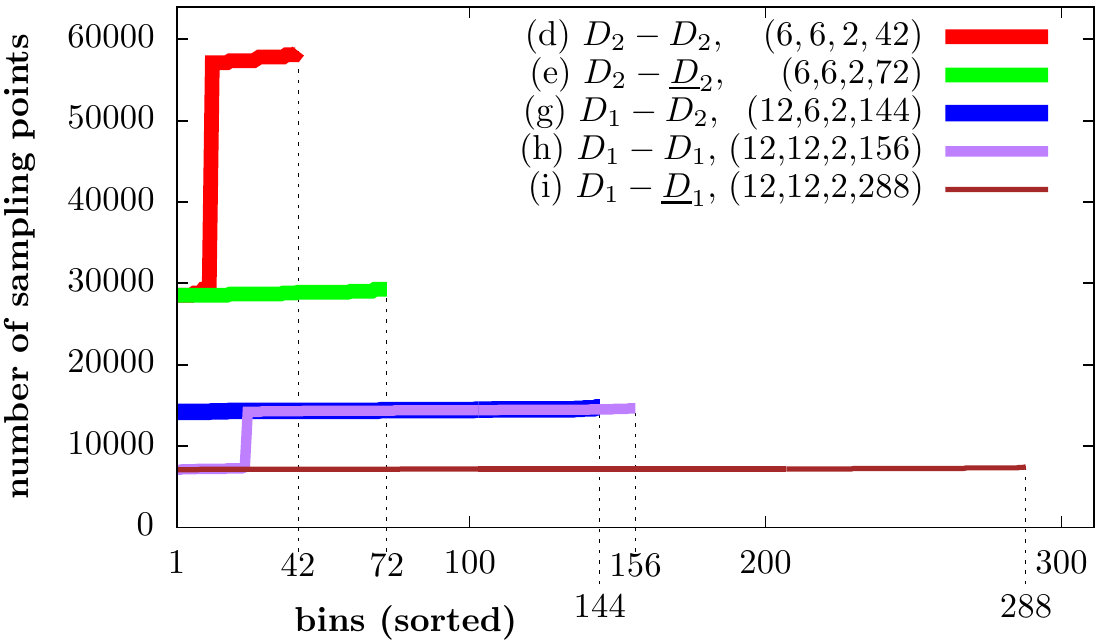}
 \end{center}
 \caption{ Sampling point distribution along the combined bins of various sets of orientation dependent reaction coordinates (defined in Table \ref{tab:2d3d} row  (d), (e), (g), (h) and (i) through the mapping functions $\bar{q}_2(\theta_{\rm A},\theta_{\rm B})$, $\bar{q}_3(\theta_{\rm A},\theta_{\rm B})$, $\bar{q}_4(\theta_{\rm A},\theta_{\rm B})$).
 Bins are sorted according to their sampling rate in ascending order.
 Underlying sampling of the angles $\theta_{\rm A}$ and $\theta_{\rm B}$ is given through  Eqs. \eqref{eqn:samp2da}-\eqref{eqn:samp2db}.
The brackets have the following code:
  $(\text{bins}(q_2),\text{bins}(q_3),\text{bins}(q_4), \text{total number of combined bins})$.
  The total number of combined bins for (d) and (h) is not equal to the product $\text{bins}(q_2) \times \dots \times \text{bins}(q_4)$ since the following holds $\text{bin}(q_2)\le \text{bin}(q_3)$ (see column 3 in Table \ref{tab:2d3d}).
  The black dotted line marks the bin number with the highest sampling rate.
}
 \label{fig:rcsamplingdist2d}
\end{figure}
\begin{figure}[htb]
\begin{center}
 \includegraphics[width=12cm]{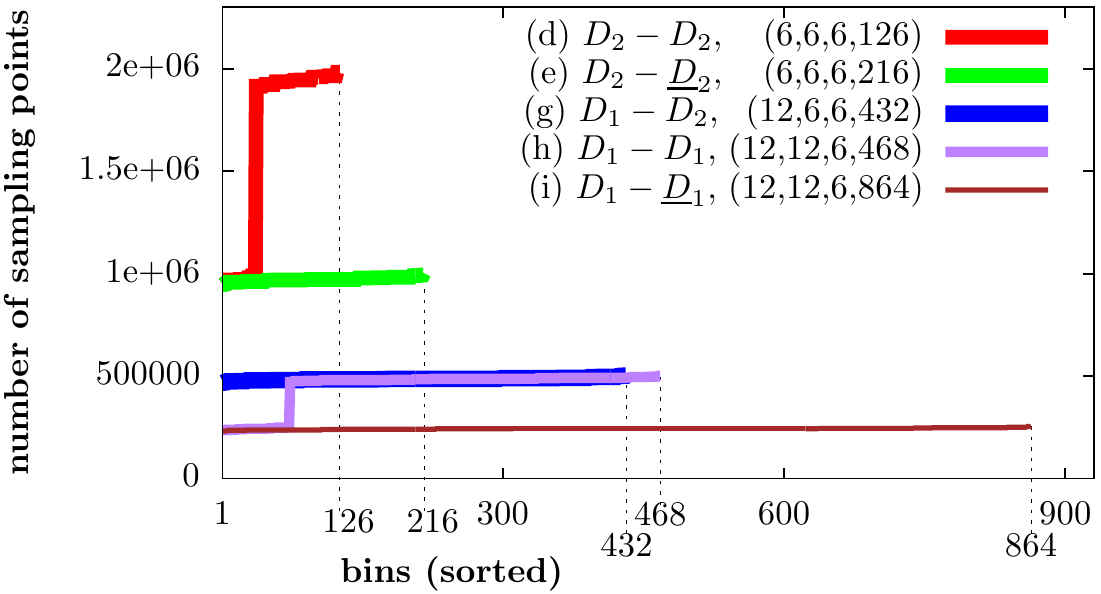}
 \end{center}
 \caption{ 
Sampling point distribution along the combined bins of various sets of orientation dependent reaction coordinates (defined in Table \ref{tab:2d3d} row  (d), (e), (g), (h) and (i) through the mapping functions $\bar{q}_2(\theta_{\rm A},\theta_{\rm B},\phi)$, $\bar{q}_3(\theta_{\rm A},\theta_{\rm B},\phi)$, $\bar{q}_4(\theta_{\rm A},\theta_{\rm B},\phi)$).
 Bins are sorted according to their sampling rate in ascending order.
 Underlying sampling of the angles $\theta_{\rm A}$, $\theta_{\rm B}$ and $\phi$ is given through  Eqs. \eqref{eqn:samp3da}-\eqref{eqn:samp3dc}.
The brackets have the following code:
  $(\text{bins}(q_2),\text{bins}(q_3),\text{bins}(q_4), \text{total number of combined bins})$.
  The total number of combined bins for (d) and (h) is not equal to the product $\text{bins}(q_2) \times \dots \times \text{bins}(q_4)$ since the following holds $\text{bin}(q_2)\le \text{bin}(q_3)$ (see column 4 in Table \ref{tab:2d3d}).
  The black dotted line marks the bin number with the highest sampling rate.
 }
 \label{fig:rcsamplingdist3d}
\end{figure}

}

\section{\label{sec:Computational methodology}Computational methodology}

{
So far, we have presented a collection of reaction coordinates for the coarse-grained description of various molecular dimers.
The current section reviews a computational methodology for obtaining an effective pair potential using the direct Boltzmann inversion principle~\cite{Tschoep1998,Tschoep1998b} as presented in a previous work \cite{Heinemann2014}.
In contrast to this previous work, our upcoming definition for the effective potential is quite general and thus compatible with all our here developed sets of reaction coordinates.
Also, we here explicitly show how the free energy of a single molecule enters the effective pair potential.
The latter fact might be interesting when coarse-graining monatomic molecules, since the single molecule free energy of those molecules is independent from the molecular species.

As we pointed out in the beginning, this effective pair potential represents one among other effective potentials and is desired for systems of large rather rigid molecules at (preferably) low densities.
Without loss of generality, we can consider for the upcoming definition of the effective pair potential the microscopic system as atomistically detailed, where each atom is characterized by its position, i.e. classical.
The atom-atom interactions are modeled through classical force-fields, representing e.g. chemical bonds or Coulomb interactions among atoms.
}

\subsection{\label{sec:Definition of the effective pair potential}Definition of the effective pair potential}

{
Many methods in the literature for calculating an effective pair potential deal more or less directly with the approximation of the mean force among coarse-grained sites (force matching schemes~\cite{Loewen1993,Ercolessi1994,Izvekov2005,Izvekov2005jcp}, iterative Boltzmann inversion method~\cite{Soper1996,MullerPlathe2002}, inverse Monte Carlo~\cite{Lyubartsev1995,Lyubartsev1997}, relative entropy method~\cite{Shell2008}, conditional reversible work method~\cite{Brini2011}, ...). 
A review concerning these methods is given in Ref. \onlinecite{Kalligiannaki2016}.
We here propose the direct Boltzmann inversion method~\cite{Tschoep1998,Tschoep1998b} (no iterations required), where the resulting effective pair potential is dependent on temperature but independent from the surrounding particle environment, i.e. independent from the system's
molecule density. The potential that is obtained using this method approximates the (density-dependent) potential of mean force among two molecules in the limit of large inter-molecule separations or low densities.
Besides many-particle effects like induction or conformational changes within the considered pair of molecules induced by surrounding molecules which cannot be considered, this type of effective pair potential has
the advantage that just two molecules have to be simulated.
In systems of rather stiff molecules this type of potential produces convenient results~\cite{Heinemann2014,Heinemann2015}.

In the following, we motivate the former described effective pair potential } by first considering probability distribution functions on different levels of detail covering the microscopic and the coarse-grained level of detail.
The main principle is to identify the microscopic probability distribution function $P_{\rm mic}$ with the coarse-grained probability distribution function $P_{\rm CG}$, which are both functions of
the considered reaction coordinates $q^M=q_1, \dots, q_M$, i.e.,
\begin{align}
 P_{\rm mic}(q^M)\equiv P_{\rm CG}(q^M)\text{.}
 \label{eqn:ansatz}
\end{align}
As we later see this approach will lead to the same definition as in a previous work~\cite{Heinemann2014}, where a rewriting of the microscopic partition sum into an integration over reaction coordinates with an effective Boltzmann weight was performed.
The microscopic probability distribution as a function of $M$ considered reaction coordinates is given by
\begin{align}
 P_{\rm mic}(q^M)= \left< \prod_{i=1}^M\delta(q_i-\tilde{q}_i(\mathbf{r}^N)) \right>_{\rm mic}\text{,}
 \label{eqn:Pmic}
\end{align}
where $\tilde{q}_i$ denotes the $i$-th mapping function that projects an atomic configuration, given through the $N$ atomic positions $\mathbf{r}^N$, on a value of the $i$-th reaction coordinate.
The brackets in Eq. \eqref{eqn:Pmic} are defined through
\begin{align}
 \left<\dots\right>_{\rm mic}=\frac{1}{\alpha} \int_{V} \mathrm{d}\mathbf{r}_1\dots\int_{V} \mathrm{d}\mathbf{r}_N       \, \frac{e^{-\frac{1}{k_{\rm B} T}\, E_{\rm pot}(\mathbf{r}^N)}}{Z} \dots
 \label{eqn:ensembleavmic}
\end{align}
and denote the canonical ensemble average in the microscopic system with $\alpha$ being a constant with a dimension of volume to the power of $N$. 
Moreover, $\alpha$ incorporates also the kinetic part of the ensemble average while $k_{\rm B}$ denotes the Boltzmann constant.
Further, $Z=Z(N,V,T)$ represents the partition function as a function of the particle number $N$, volume $V$ and temperature $T$, while $E_{\rm pot}(\mathbf{r}^N)$ represents the potential energy of the 
atomic configuration $\mathbf{r}^N$.
The coarse-grained probability distribution from Eq. \eqref{eqn:ansatz} is given through
\begin{align}
 P_{\rm CG}(q^M)= \left< \prod_{i=1}^M\delta(q_i-\mathfrak{q}_i) \right>_{\rm CG}\text{,}
 \label{eqn:PCG}
\end{align}
with the canonical ensemble average in the coarse-grained system
\begin{align}
  \left<\dots \right>_{\rm CG}= \frac{1}{\mathfrak{a}} \int_{ \mathfrak{Q}_1} \!\!\mathrm{d} \mathfrak{q}_1\dots\int_{\mathfrak{Q}_M}\!\! \mathrm{d} \mathfrak{q}_M \, J(\mathfrak{q}^M)\cdot     \, \frac{e^{-\frac{1}{k_{\rm B} T}\, U_{\rm eff}(\mathfrak{q}^M)}}{\mathfrak{Z}}\dots \;\text{,}
  \label{eqn:ensembleavcg}
\end{align}
where each $\mathfrak{Q}_i$ denotes the value set of the $i$-th reaction coordinate.
By defining the expressions
\begin{subequations}
 \allowdisplaybreaks
\begin{align}
J(\mathfrak{q}^M) &=\frac{1}{\alpha} \int_{V} \mathrm{d}\mathbf{r}_1\dots\int_{V} \mathrm{d}\mathbf{r}_N       \,  \prod_i\delta(\mathfrak{q}_i-\tilde{q}_i(\mathbf{r}^N)) 
\label{eqn:J}
\\
\mathfrak{Z}&=Z
\end{align}
\end{subequations}
with $J$ counting the microstates belonging to a macrostate (configuration space {density} of $q^M$),
we finally arrive at a definition for the effective pair potential, yielding
\begin{align}
 U_{\rm eff}(q^M)&=-k_{\rm B} T\ln \left[  \frac{Z \cdot\mathfrak{a}}{ J(q^M)}    \;P_{\rm mic}(q^M)   \right]\text{,}
 \label{eqn:ueff}
\end{align}
where only $\mathfrak{a}$ is not explicitly defined.
At this point it is important to emphasize that the previously defined probability functions (see Eqs. \eqref{eqn:Pmic} and \eqref{eqn:PCG}) are in general not dimensionless.
Regarding Eq. \eqref{eqn:ueff}, it can be shown \cite{Heinemannphd2016} that this definition is compatible with the idea of Kirkwood's potential of mean force \cite{Kirkwood1935}, where the expression
in the brackets of Eq. \eqref{eqn:ueff} is identified with a pair correlation function.
However, we here consider internal degrees of freedom and only two molecules.
Similar definitions were already presented\cite{Sippl1990,Buchete2004,Fan2011,Philips2013,Liu2015,Poier2015}.

We next develop an expression for $\mathfrak{a}$.
A common requirement for a pair potential is that it vanishes for large inter-molecular separations, i.e.,
\begin{align}
 U_{\rm eff}(q^M)\xrightarrow{R \rightarrow \infty} 0\text{.}
\end{align}
This is equivalent to the case that the argument of the logarithm in Eq.~\eqref{eqn:ueff} approaches unity for infinite inter-molecular separations.
By considering the following function 
\begin{align}
  f(q^M)=\frac{P_{\rm mic}(q^M)\,Z}{J(q^M)}
  \label{eqn:f}
\end{align}
it immediately follows
\begin{align}
U_{\rm eff}(q^M)\xrightarrow{R \rightarrow \infty} 0 \Rightarrow f(q^M)\xrightarrow{R \rightarrow \infty} 1/\mathfrak{a}\text{.}
\label{eqn:limit}
\end{align}
By using the definitions in Eq. \eqref{eqn:Pmic}, \eqref{eqn:ensembleavmic} and \eqref{eqn:J}, Eq.~\eqref{eqn:f} can be written as
\begin{align}
f(q^M)=\frac{ \int_{V} \mathrm{d}\mathbf{r}_1\dots\int_{V} \mathrm{d}\mathbf{r}_N       \,  \prod_i\delta(q_i-\tilde{q}_i(\mathbf{r}^N))\,\exp\left(-\frac{1}{k_{\rm B} T}\, E_{\rm pot}(\mathbf{r}^N)\right)  }{ \int_{V} \mathrm{d}\mathbf{r}_1\dots\int_{V} \mathrm{d}\mathbf{r}_N       \, \prod_i\delta(q_i-\tilde{q}_i(\mathbf{r}^N)) }\text{.}
\label{eqn:fnewer}
\end{align}
Hereby, the expression $\exp\left(-\frac{1}{k_{\rm B} T}\, E_{\rm pot}(\mathbf{r}^N)\right)$ is averaged over all internal coordinates belonging to the reaction coordinates $q^M$.
At this point it seems important to emphasize that the reaction coordinates are chosen to constrain only the relative arrangement of the group of points $\mathbf{r}_1$,\dots,$\mathbf{r}_{N_{\rm A}}$ belonging to molecule A
and the group of points $\mathbf{r}_{N_{\rm A}+1}$,\dots,$\mathbf{r}_N$ belonging to molecule B.
However, when we consider large inter-molecular separations, the potential energy splits separately into the molecular parts, i.e.,
\begin{align}
 E_{\rm pot}(\mathbf{r}^N)\xrightarrow{R \rightarrow \infty} \epsilon_{\rm pot}(\mathbf{r}_1, \dots, \mathbf{r}_{N_{\rm A}}) + \epsilon_{\rm pot}(\mathbf{r}_{N_A+1}, \dots, \mathbf{r}_N)\text{.}
\end{align}
{Consequently, $\exp\left(-\frac{1}{k_{\rm B} T}\, E_{\rm pot}(\mathbf{r}^N)\right)$ from Eq. \eqref{eqn:fnewer} does not depend on the relative molecular distance and orientation any longer and it is thus equivalent to define 
$f$ via an average of $\exp\left(-\frac{1}{k_{\rm B} T}\, E_{\rm pot}(\mathbf{r}^N)\right)$ over the entire configuration space.}
The function $f$ thus fulfills the following factorization
\begin{multline}
 f(q^M)\xrightarrow{R \rightarrow \infty} \left(\frac{1}{V^{N_{\rm A}}}\int_{V} \mathrm{d}\mathbf{r}_1\dots\int_{V} \mathrm{d}\mathbf{r}_{N_{\rm A}}      \, e^{-\frac{1}{k_{\rm B} T}\, \epsilon_{\rm pot}(\mathbf{r}_1, \dots, \mathbf{r}_{N_{\rm A}})} \right)
 \\
 \times\left(\frac{1}{V^{N-N_{\rm A}}}\int_{V} \mathrm{d}\mathbf{r}_{N_{\rm A}+1}\dots\int_{V} \mathrm{d}\mathbf{r}_{N}       \, e^{-\frac{1}{k_{\rm B} T}\, \epsilon_{\rm pot}(\mathbf{r}_{N_{{\rm A}+1}}, \dots, \mathbf{r}_{N})}  \right)\text{.}
 \label{eqn:f-factorization}
\end{multline}
Each factor can be written in terms of the temperature de\-pen\-dent free energy $F^{\rm A / B}$ of molecule A or B, respectively.
As a result, we arrive at the following limit
\begin{align}
 f(q^M)\xrightarrow{R \rightarrow \infty}\left(\frac{\alpha}{V^N}\right)^2 \cdot \exp\left(-\frac{1}{k T} \left( F^{\rm A}(T)+F^{\rm B}(T) \right)\right)\text{.}
\end{align}
By using Eq. \eqref{eqn:limit}, we obtain a temperature dependent definition for $\mathfrak{a}$, namely
\begin{align}
  \mathfrak{a}=\mathfrak{a}(T)&=\lim_{R\rightarrow \infty} \frac{1}{f(q^M)}\\
 &=\left(\frac{V^N}{\alpha}\right)^2 \cdot \exp\left(+\frac{1}{k T} \left( F^{\rm A}(T)+F^{\rm B}(T) \right)\right)
 \text{.}
\end{align}
At this point it is worthwhile to mention that our theory in this section foots on the canonical ensemble.
For other statistical ensembles the ensemble averages defined in Eq. \eqref{eqn:ensembleavmic} and \eqref{eqn:ensembleavcg} have to be modified accordingly.

\subsection{\label{sec:General algorithm}General algorithm}

The following algorithm acts as a recipe for calculating effective molecular pair potentials {at a specific temperature}.
In case of pronounced potential barriers or low temperatures, modifications towards rare-event sampling methods as presented in Appendix \ref{sec:Rare-event sampling methods} have to be performed.
{
We already used this guideline in a previous investigation for determining an angle-resolved effective potential among coronene molecules~\cite{Heinemann2014}.
In that investigation, however, the reaction coordinates were differently chosen as the ones presented here to be fully compatible with the Berne-Pechukas~\cite{Berne1972} or Gay-Berne potential~\cite{Gay1981}.
Since we here propose the direct use of tabulated interactions there is no further need to consider compatibility of reaction coordinates with potential parameters of established models.

The fundamental steps can be summarized as follows:
}

\begin{enumerate}
 \item create atomic trajectories / configurations with standard packages (e.g. GROMACS, LAMMPS, ESPREesSo\dots) in the desired ensemble
 \item project each configuration $\mathbf{r}^N$ onto the position vectors $\mathbf{R}_{\rm A}$, $\mathbf{R}_{\rm B}$ (e.g., using the center of masses) and orientation vectors $\hat{\mathbf{u}}_{\rm A}$, $\hat{\mathbf{u}}_{\rm A}$
 (e.g., using for each molecule the axis with the largest moment of inertia)
 \item calculate $R$, $\theta_{\rm A}$ and $\theta_{\rm B}$ defined through Eqs. \eqref{eqn:rca} - \eqref{eqn:rcc} and additionally $\phi$ for the three-dimensional case as defined through Eq. \eqref{eqn:rcphi}
 \item use mapping functions from Table \ref{tab:2d3d} to create reaction coordinates
 \label{rc table}
 \item discretize the table from step \ref{rc table}
 \label{discretizing step}
 \item sort resulting table, e.g. with the Linux ``sort'' command on multiple fields
 $$\text{sort}\, \text{-k1n,1}\, \text{-k2n,2}\, ... \,\textit{file\_in} > \textit{file\_out}$$
 \item create a histogram (probability distribution function) by counting equal (neighboring) rows in \textit{file\_out}
 \item tabulate (if necessary) the $J$-function (Eq. \eqref{eqn:J}) by using the following formula
 \\
 for the two-dimensional case:
 \begin{align}
J(q^M)\sim const\cdot \int_0^{2\pi}\!\! \mathrm{d}\theta_{\rm A} \int_0^{2\pi}\!\! \mathrm{d}\theta_{\rm B} \,\mathfrak{F}\left((q^-)^M,(q^+)^M\right)
\end{align}
 for the three-dimensional case:
 \begin{align}
J(q^M)\sim const\cdot\int_0^{\pi} \mathrm{d}\theta_{\rm A}\,\sin \theta_{\rm A}\,\int_0^{\pi} \mathrm{d}\theta_{\rm B}\, \sin\theta_{\rm B}   \int_0^{2\pi} \mathrm{d}\phi  \,\mathfrak{F}\left((q^-)^M,(q^+)^M\right)
\end{align}
with the function
$$\mathfrak{F}\left((q^-)^M,(q^+)^M\right)=\begin{cases}1&, \bar{q}_i(\theta_{\rm A},\dots) \in \left[q_i^-,q_i^+\right) \forall i\\0&, \text{else} \end{cases}$$
yielding unity if the set of reaction coordinate values lies in the considered range $[(q^-)^M,(q^+)^M)$ around the $q^M$, otherwise zero.
 \item calculate the effective potential up to a shift using Eq. \eqref{eqn:ueff}; the shift can be determined at large inter-molecular distances
\end{enumerate}

\section{\label{sec:Conclusion}Conclusion}

We have presented a strategy to determine a tabulated effective pair potential between two particles (preferably molecules) of the same or different species with various symmetries.
For this purpose, our study focuses on molecules with mainly spherical, polar or uniaxial symmetry in two- and three-dimensional systems for which we 
developed appropriate sets of reaction coordinates (see Table \ref{tab:2d3d}).
These coordinates are chosen to take the different particle symmetries into account while avoiding any redundancies in the coarse-grained particle descriptions.
The {numerical} table describing the interactions is thus minimal in length, { and requires a minimal memory storage as well as look-up time accordingly}.
Depending on the molecules' composition, shape and charge distribution, an effective pair potential taking the anisotropies (rather than only molecular distance dependence) into account, may significantly
improve the coarse-grained description while still neglecting irrelevant degrees of freedom (e.g., atomic positions within a molecule).

The numerical method (see Sec. \ref{sec:Numerical evidence of the optimal choice of reaction coordinates}) - we used to check the optimal choice of reaction coordinates - can be applied to other types of molecular pairs with more complex symmetry where a proof of the optimality is far from trivial.
{A slight drawback of our choice of non-redundant reaction coordinates might be the rather complex associated mathematical expressions,
which can be inadequate for the direct use as parameters in a model potential. Furthermore some of these coordinates do not seem to be intuitive at a first glance.
For a more intuitive representation, a redundant set of coordinates might in some cases be more appropriate.
}
A primary factor in determining the effective pair potential through the direct Boltzmann inversion is the unbiased probability distribution, which is determined through the sampling of the reaction coordinates.
Here, we {also refer to} two methods to overcome potential barriers that hamper the sampling in Appendix \ref{sec:Rare-event sampling methods}.
These methods improve sampling along the inter-molecular distance. {Overcoming also} orientational barriers could be important for strongly anisotropic molecules.
A solution for the latter issue is given by Ba\-ba\-di {\it{et al.}} \cite{Babadi2006}, who used a dynamics, which constrained the primary molecular orientation (see also Ref. \onlinecite{Hess2003}) to create potential paths for further pa\-rametri\-zation.
However, determining the entire potential with orientational constraints / restraints is quite costly from a computational prospective.

The overall strategy of this work can be applied to various types of natural molecules, and also artificial (man-made) particles with high rotational mobility (when using the presented sampling methods in Appendix \ref{sec:Rare-event sampling methods}).  
This strategy has already been applied for coronene molecules \cite{Heinemann2014} (leading to a speed-up of factor 3 to 8 per CPU core compared to the atomic system) and ring polymers \cite{Poier2015}, but with sets of reaction coordinates fully compatible with the Gay-Berne
potential \cite{Gay1981} parameters.
A further interesting benchmark system is given by molecules from the air as partly depicted in Fig. \ref{fig:types0}.
These molecules are simple enough to allow {\emph {ab-initio}} simulations \cite{Car1985,Marx2000}, i.e. first principles calculations combined with molecular dynamics, as an elaborate method for the microscopic simulations.
Also, we would like to point out that we are not restricted to any particular length scale in this investigation.
A real colloidal system provides {also} a large-scale application for our sets of reaction coordinates.
Video tracking the particles \cite{Muller2014} - perhaps with orientation - allows the calculation of the distribution functions and thus the pair potential (for two isolated colloids) or the potential of mean force in a dense system with many colloids (the {so developed environmental dependent} pair potential can be obtained using integral equation schemes \cite{Li2005} or iterative Boltzmann inversion schemes~\cite{Soper1996,MullerPlathe2002}).
Having determined the effective pair potential, it is then possible to simulate thousands of molecules {\emph {in silico}}.
An interesting field for further research is to define sets of reaction coordinates involving biaxial particles, such as diindenoperylene (DIP) mo\-le\-cules.
Besides elaborate particle-particle (e.g., molecule-molecule) descriptions, a quite interesting challenge is to find adequate sets of reaction coordinates for describing molecules relative to an inhomogeneous surface (e.g., striped surfaces \cite{DellaSala2011,Palczynski2014b,Kleppmann2015}).

\section{Acknowledgement}

This work was partly supported by the Deutsche Forschungsgemeinschaft (DFG) within the framework of the CRC 951
(Project No. A7).
Furthermore, we thank Doctor K. Palczynski and Professor J. Dzubiella for fruitful discussions.

\appendix
{
\section{\label{sec:Exemplary proof of f being bijective}Exemplary proof of $f$ being bijective}

In the following, we proof that the function $f$ as defined in \eqref{eqn:f1} and \eqref{eqn:f2}, with the explicit reaction coordinates given in Table \ref{tab:2d3d}, is bijective for the $D_1-D_2$ dimer type (see Fig. \ref{fig:rcpics} (g)) in two dimensions.
The proofs for the remaining non-trivial dimer types (i.e. $D_1-D_1$, $D_1-\underline{D}_1$, $D_2-D_2$, $D_2-\underline{D}_2$) can be done in a similar fashion.
First, we proof $f$ being injective as defined in Eq. \eqref{eqn:injective}.
For this purpose, we explore all parametrizations belonging to an equal set of reaction coordinate values by analyzing each vector component in the equation $f(R,\theta_{\rm A},\theta_{\rm B}) = f(R',\theta_{\rm A}',\theta_{\rm B}')$: 
\begin{enumerate}
 \item $\bar{q}_1(R,\theta_{\rm A},\theta_{\rm B})=R^2\stackrel{!}{=} (R')^2=\bar{q}_1(R',\theta_{\rm A}',\theta_{\rm B}')$
 \\  $\hookrightarrow$ $R'=R$ for $R,R'\in [0,\infty)$
\item $\bar{q}_2(R,\theta_{\rm A},\theta_{\rm B})=\frac{1}{\pi}\arccos\cos(\theta_{\rm A})\stackrel{!}{=} \frac{1}{\pi}\arccos\cos(\theta_{\rm A}')=\bar{q}_2(R',\theta_{\rm A}',\theta_{\rm B}')$ 
  \\  $\hookrightarrow$ $\theta_{\rm A}'=\theta_{\rm A}$ and $\theta_{\rm A}'=-\theta_{\rm A}$ for $\theta_{\rm A},\theta_{\rm A}'\in [-\pi,\pi]$
  \item $\bar{q}_3(R,\theta_{\rm A},\theta_{\rm B}) = \frac{2}{\pi}\arccos|\cos(\theta_{\rm B})| \stackrel{!}{=} \frac{2}{\pi}\arccos|\cos(\theta_{\rm B}')| = \bar{q}_3(R',\theta_{\rm A}',\theta_{\rm B}')$
  \\
 $\hookrightarrow$  $\theta_{\rm B}'=\theta_{\rm B}, -\theta_{\rm B}, \pi-|\theta_{\rm B}|, -\pi+|\theta_{\rm B}|$  for $\theta_{\rm B},\theta_{\rm B}'\in [-\pi,\pi]$
  \item $\bar{q}_4(\theta_{\rm A},\theta_{\rm B})=\text{sgn}_{+}(\sin(\theta_{\rm A})\cdot\sin(2\theta_{\rm B}))  \stackrel{!}{=}  \text{sgn}_{+}(\sin(\theta_{\rm A}')\cdot\sin(2\theta_{\rm B}')) =\bar{q}_4(\theta_{\rm A}',\theta_{\rm B}')   $
\end{enumerate}
The last equation reduces the solutions to
\begin{align}
\label{eqn:solutiond1d2dimer2d}
 (R',\theta_{\rm A}',\theta_{\rm B}')\in
 \left\{ \begin{array}{c}(R,\theta_{\rm A},\theta_{\rm B}),\\
 (R,-\theta_{\rm A},-\theta_{\rm B}),\\
 (R,\theta_{\rm A}, \text{sgn}_+(\theta_{\rm B}) \cdot (-\pi+|\theta_{\rm B}|) + \text{sgn}_+(-\theta_{\rm B}) \cdot ( \pi - |\theta_{\rm B}| ) ), \\
 (R,-\theta_{\rm A}, \text{sgn}_+(-\theta_{\rm B}) \cdot (-\pi+|\theta_{\rm B}|) + \text{sgn}_+(\theta_{\rm B}) \cdot ( \pi - |\theta_{\rm B}| ) )
 \end{array}
 \right\}
\end{align}
To proof that all the possible parametrizations form an equivalence class with respect to the symmetry functions $\tilde{F}_{\rm B}$, $\tilde{\chi}$, we need to show that combinations of the latter
functions are able to transform each parametrization into one another (see \eqref{eqn:2dcgstate}). Since these satisfy  
$$\tilde{F}_{\rm B}\circ{} \tilde{F}_{\rm B}=\tilde{I}_d,  \;[\tilde{F}_{\rm B},\tilde{\chi}]=0,\; \tilde{\chi}\circ{}\tilde{\chi}=\tilde{I}_d\; \text{with} \; \tilde{I}_d=\alpha \circ{} \text{id},$$
we only have to show that for all parametrizations $(R',\theta_{\rm A}',\theta_{\rm B}')$ from \eqref{eqn:solutiond1d2dimer2d}, $\exists \tilde{S} \in \left\{\tilde{I}_d, \tilde{\chi}, \tilde{F}_{\rm B}, \tilde{F}_{\rm B} \circ{} \tilde{\chi} \right\}$ with $\tilde{S}(R',\theta_{\rm A}',\theta_{\rm B}')=(R,\theta_{\rm A},\theta_{\rm B})$, which is trivial.
For the proofs of the remaining dimer types, we can use the other relations defined in Eq. \eqref{eqn:relationsa}--\eqref{eqn:relationsc}.

We next proof $f$ being surjective as defined in Eq. \eqref{eqn:surjective}.
The corresponding functions $\bar{q}_1=\bar{q}_1(R)$, $\bar{q}_2=\bar{q}_2(\theta_{\rm A})$ and $\bar{q}_3=\bar{q}_3(\theta_{\rm B})$ are obviously surjective and independent from each other.
The remaining surjective reaction coordinate $\bar{q}_4$, however, depends on $\theta_{\rm A}$ and $\theta_{\rm B}$ appearing in $\bar{q}_2$ and $\bar{q}_3$ and can only take the values $0$ and $1$.
Thus, we know that
\begin{multline}
\exists X\in \{0,1\} : \forall q_1,q_2, q_3 \;\text{with}\; (q_1,q_2,q_3,X)^{T} \in \mathfrak{Q} \, \exists \, R, \theta_{\rm A}, \theta_{\rm B} \\ \text{with} \;f(R, \theta_{\rm A}, \theta_{\rm B})=(q_1,q_2,q_3,X)^{T}.
\label{eqn:surjective2}
\end{multline}
To proof Eq. \eqref{eqn:surjective}, we still have to show that Eq. \eqref{eqn:surjective2} holds for all $X\in \{0,1\}$.
By choosing $(R', \theta_{\rm A}', \theta_{\rm B}') =(R, -\theta_{\rm A}, \theta_{\rm B})$ it follows that $f(R', \theta_{\rm A}', \theta_{\rm B}')=(q_1,q_2,q_3,1-X)^{T}.$
Accordingly, the function $f$ is surjective and due to its injective character also bijiective.

}

\section{\label{sec:Rare-event sampling methods}Rare-event sampling methods}

The calculation of the probability distribution defined by Eq. \eqref{eqn:Pmic}, which is required to calculate the effective pair potential (see Eq. \eqref{eqn:ueff}),
is generally quite challenging since the sampling of the underlying molecular dynamics or Monte Carlo simulations might be hampered by potential barriers.
This barriers can become insurmountable at low temperatures and this might lead to the case that the two molecules are glued together during the whole simulation.
There are various methods improving this drawback like steered dynamics (SD) \cite{Straatsma1992}, umbrella sampling (US) \cite{Torrie1974,Torrie1977}, metadynamics~\cite{Laio2002}, parallel tempering \cite{Chodera2007} and further derivatives like adaptive US \cite{Mezei1987} or umbrella integration \cite{Kastner2005,Kastner2009}.
We here briefly summarize the main idea of SD and US for improving distance dependent sampling when performing distance and orientation dependent coarse-graining in the canonical ensemble.
A comparison between both methods for discotic mole\-cules was already performed in Ref. \onlinecite{Heinemann2014}.

In the SD method the two mole\-cules are pulled apart and the work applied to this system is integrated with respect to the pull coordinate $q_1$ to an effective pair potential (see thermodynamic integration \cite{Kirkwood1935} or Ref. \onlinecite{Park2003} for a non-equilibrium approach).
The equilibrium probability distribution $P(q_1)$ can then be recovered by inverting Eq. \eqref{eqn:ueff} using $M=1$.
This method can be extended towards the distance and orientation dependent case by orientational constraints \cite{Hess2003}.
Consequently, for each orientation an extra pull simulation has to be performed.
Alternatively, for calculating the entire histogram (distance and orientation dependent, i.e., $P(q^M)$), one can use the following factorization of the probability distribution into a pure distance dependent part
(we assume for the first reaction coordinate $q_1=R$, i.e. a pure distance dependence such as the center of mass distance) and an orientation dependent part as pointed out in Ref. \onlinecite{Heinemann2014}, that is
\begin{align}
 P(q^M)=P(R)\cdot \frac{P(q^M)}{P(R)}\text{.}
\end{align}
The underlying idea behind the factorization is that the molecules are pulled apart on a much larger time scale than those of the orientational degrees of freedom.
We assume that in the vicinity of each distance $R$ the corresponding orientational histogram $P(q^M|q_1=R)= \frac{P(q^M)}{P(R)}$ can  be tracked.
Finally, the combined histogram $P(q^M)$ leads to the distance and orientation dependent effective potential $U_{\rm eff}$ as defined in Eq. \eqref{eqn:ueff}.

In the US method many simulations or so-called ``windows'' are performed where the system of both molecules is biased by an additional potential.
Usually a spring-like potential between the center of masses - denoted with $R$ - is used.
The spring potential of the $k$-th simulation is often chosen harmonic, i.e.,
\begin{align}
 V_k(R)=K/2\cdot (R-R^{\rm eq}_k)^2\text{.}
\end{align}
By varying the spring constant $K$, the equilibrium length $R^{\rm eq}_k$ {but also $T_i$ (temperature in the i-th umbrella window)} one can achieve a great overlap of the biased probability distribution functions $P_k^{\text{bias}}(q^M)$ and $P_{k+1}^{\text{bias}}(q^M)$ of neighboring umbrella windows.
The overlap is necessary when creating an unbiased histogram function $P(q^M)$ for the whole range of $R$ by using the weighted histogram analysis method (WHAM)  \cite{Kumar1992,Roux1995}.
In this method, the following set of equations (considering one temperature $T_i=T$ for all simulations):
\begin{subequations}\label{eqn:WHAMoriginal}
\begin{align}
 \label{eqn:WHAMoriginala}
P(R)&=\sum_{k=1}^{N_w} \gamma_k(R) \, P_k^{\text{bias}}(R)\\
\label{eqn:WHAMoriginalb}
\gamma_k(R)&=\frac{n_k} { \sum_{i=1}^{N_w} n_i \, e^{-\frac{1}{k_{\rm B} T} (V_i(R)-F_i)}  }\\
\label{eqn:WHAMoriginalc}
F_k&=-k_{\rm B} T \ln\left[\int_{\mathfrak{Q}_1}\!\!\! \mathrm{d}R \,e^{-\frac{1}{k_{\rm B} T} V_k(R)} P(R) \right]
\end{align}
\end{subequations}
is solved self-consistently, where $\gamma_k(R)$ marks the weight of each histogram, $N_w$ the number of umbrella windows, $n_k$ the number of sampling points in umbrella window $k$ and $F_k$ a free energy constant.
The total unbiased histogram function can then be determined using previous weights \cite{Heinemann2014}, that is
\begin{align}
 P(q^M)&=\sum_{k=1}^{N_w} \gamma_k(R) \, P_k^{\text{bias}}(q^M)\text{.}
\end{align}
Furthermore, we want to point out that both of the presented methods can be simply rewritten for different definitions of $q_1$ with higher orders in the distance dependence as defined in Table \ref{tab:2d3d}.

%
%
%
%
%
%




\begin{thebibliography}{73}%
\makeatletter
\providecommand \@ifxundefined [1]{%
 \@ifx{#1\undefined}
}%
\providecommand \@ifnum [1]{%
 \ifnum #1\expandafter \@firstoftwo
 \else \expandafter \@secondoftwo
 \fi
}%
\providecommand \@ifx [1]{%
 \ifx #1\expandafter \@firstoftwo
 \else \expandafter \@secondoftwo
 \fi
}%
\providecommand \natexlab [1]{#1}%
\providecommand \enquote  [1]{``#1''}%
\providecommand \bibnamefont  [1]{#1}%
\providecommand \bibfnamefont [1]{#1}%
\providecommand \citenamefont [1]{#1}%
\providecommand \href@noop [0]{\@secondoftwo}%
\providecommand \href [0]{\begingroup \@sanitize@url \@href}%
\providecommand \@href[1]{\@@startlink{#1}\@@href}%
\providecommand \@@href[1]{\endgroup#1\@@endlink}%
\providecommand \@sanitize@url [0]{\catcode `\\12\catcode `\$12\catcode
  `\&12\catcode `\#12\catcode `\^12\catcode `\_12\catcode `\%12\relax}%
\providecommand \@@startlink[1]{}%
\providecommand \@@endlink[0]{}%
\providecommand \url  [0]{\begingroup\@sanitize@url \@url }%
\providecommand \@url [1]{\endgroup\@href {#1}{\urlprefix }}%
\providecommand \urlprefix  [0]{URL }%
\providecommand \Eprint [0]{\href }%
\providecommand \doibase [0]{http://dx.doi.org/}%
\providecommand \selectlanguage [0]{\@gobble}%
\providecommand \bibinfo  [0]{\@secondoftwo}%
\providecommand \bibfield  [0]{\@secondoftwo}%
\providecommand \translation [1]{[#1]}%
\providecommand \BibitemOpen [0]{}%
\providecommand \bibitemStop [0]{}%
\providecommand \bibitemNoStop [0]{.\EOS\space}%
\providecommand \EOS [0]{\spacefactor3000\relax}%
\providecommand \BibitemShut  [1]{\csname bibitem#1\endcsname}%
\let\auto@bib@innerbib\@empty
\bibitem [{\citenamefont {Stockmayer}(1941)}]{Stockmayer1941}%
  \BibitemOpen
  \bibfield  {author} {\bibinfo {author} {\bibfnamefont {W.~H.}\ \bibnamefont
  {Stockmayer}},\ }\href {\doibase http://dx.doi.org/10.1063/1.1750922}
  {\bibfield  {journal} {\bibinfo  {journal} {J. Chem. Phys.}\ }\textbf
  {\bibinfo {volume} {9}},\ \bibinfo {pages} {398} (\bibinfo {year}
  {1941})}\BibitemShut {NoStop}%
\bibitem [{\citenamefont {Berne}\ and\ \citenamefont
  {Pechukas}(1972)}]{Berne1972}%
  \BibitemOpen
  \bibfield  {author} {\bibinfo {author} {\bibfnamefont {B.~J.}\ \bibnamefont
  {Berne}}\ and\ \bibinfo {author} {\bibfnamefont {P.}~\bibnamefont
  {Pechukas}},\ }\href@noop {} {\bibfield  {journal} {\bibinfo  {journal} {J.
  Chem. Phys.}\ }\textbf {\bibinfo {volume} {56}},\ \bibinfo {pages} {4213}
  (\bibinfo {year} {1972})}\BibitemShut {NoStop}%
\bibitem [{\citenamefont {Gay}\ and\ \citenamefont {Berne}(1981)}]{Gay1981}%
  \BibitemOpen
  \bibfield  {author} {\bibinfo {author} {\bibfnamefont {J.~G.}\ \bibnamefont
  {Gay}}\ and\ \bibinfo {author} {\bibfnamefont {B.~J.}\ \bibnamefont
  {Berne}},\ }\href {\doibase 10.1063/1.441483} {\bibfield  {journal} {\bibinfo
   {journal} {J. Chem. Phys.}\ }\textbf {\bibinfo {volume} {74}},\ \bibinfo
  {pages} {3316} (\bibinfo {year} {1981})}\BibitemShut {NoStop}%
\bibitem [{\citenamefont {Cleaver}\ \emph {et~al.}(1996)\citenamefont
  {Cleaver}, \citenamefont {Care}, \citenamefont {Allen},\ and\ \citenamefont
  {Neal}}]{Cleaver1996}%
  \BibitemOpen
  \bibfield  {author} {\bibinfo {author} {\bibfnamefont {D.~J.}\ \bibnamefont
  {Cleaver}}, \bibinfo {author} {\bibfnamefont {C.~M.}\ \bibnamefont {Care}},
  \bibinfo {author} {\bibfnamefont {M.~P.}\ \bibnamefont {Allen}}, \ and\
  \bibinfo {author} {\bibfnamefont {M.~P.}\ \bibnamefont {Neal}},\ }\href
  {\doibase 10.1103/PhysRevE.54.559} {\bibfield  {journal} {\bibinfo  {journal}
  {Phys. Rev. E}\ }\textbf {\bibinfo {volume} {54}},\ \bibinfo {pages} {559}
  (\bibinfo {year} {1996})}\BibitemShut {NoStop}%
\bibitem [{\citenamefont {L\"owen}\ and\ \citenamefont
  {Kramposthuber}(1993)}]{Loewen1993}%
  \BibitemOpen
  \bibfield  {author} {\bibinfo {author} {\bibfnamefont {H.}~\bibnamefont
  {L\"owen}}\ and\ \bibinfo {author} {\bibfnamefont {G.}~\bibnamefont
  {Kramposthuber}},\ }\href {\doibase 10.1209/0295-5075/23/9/009} {\bibfield
  {journal} {\bibinfo  {journal} {Europhys. Lett.}\ }\textbf {\bibinfo {volume}
  {23}},\ \bibinfo {pages} {673} (\bibinfo {year} {1993})}\BibitemShut
  {NoStop}%
\bibitem [{\citenamefont {Ercolessi}\ and\ \citenamefont
  {Adams}(1994)}]{Ercolessi1994}%
  \BibitemOpen
  \bibfield  {author} {\bibinfo {author} {\bibfnamefont {F.}~\bibnamefont
  {Ercolessi}}\ and\ \bibinfo {author} {\bibfnamefont {J.~B.}\ \bibnamefont
  {Adams}},\ }\href {\doibase 10.1209/0295-5075/26/8/005} {\bibfield  {journal}
  {\bibinfo  {journal} {Europhys. Lett.}\ }\textbf {\bibinfo {volume} {26}},\
  \bibinfo {pages} {583} (\bibinfo {year} {1994})}\BibitemShut {NoStop}%
\bibitem [{\citenamefont {Izvekov}\ and\ \citenamefont
  {Voth}(2005{\natexlab{a}})}]{Izvekov2005}%
  \BibitemOpen
  \bibfield  {author} {\bibinfo {author} {\bibfnamefont {S.}~\bibnamefont
  {Izvekov}}\ and\ \bibinfo {author} {\bibfnamefont {G.~A.}\ \bibnamefont
  {Voth}},\ }\href {\doibase 10.1021/jp044629q} {\bibfield  {journal} {\bibinfo
   {journal} {J. Phys. Chem. B}\ }\textbf {\bibinfo {volume} {109}},\ \bibinfo
  {pages} {2469} (\bibinfo {year} {2005}{\natexlab{a}})}\BibitemShut {NoStop}%
\bibitem [{\citenamefont {Izvekov}\ and\ \citenamefont
  {Voth}(2005{\natexlab{b}})}]{Izvekov2005jcp}%
  \BibitemOpen
  \bibfield  {author} {\bibinfo {author} {\bibfnamefont {S.}~\bibnamefont
  {Izvekov}}\ and\ \bibinfo {author} {\bibfnamefont {G.~A.}\ \bibnamefont
  {Voth}},\ }\href@noop {} {\bibfield  {journal} {\bibinfo  {journal} {J. Chem.
  Phys.}\ }\textbf {\bibinfo {volume} {123}},\ \bibinfo {pages} {134105}
  (\bibinfo {year} {2005}{\natexlab{b}})}\BibitemShut {NoStop}%
\bibitem [{\citenamefont {Shell}(2008)}]{Shell2008}%
  \BibitemOpen
  \bibfield  {author} {\bibinfo {author} {\bibfnamefont {M.~S.}\ \bibnamefont
  {Shell}},\ }\href@noop {} {\bibfield  {journal} {\bibinfo  {journal} {J.
  Chem. Phys.}\ }\textbf {\bibinfo {volume} {129}},\ \bibinfo {eid} {144108}
  (\bibinfo {year} {2008})}\BibitemShut {NoStop}%
\bibitem [{\citenamefont {Brini}, \citenamefont {Marcon},\ and\ \citenamefont
  {van~der Vegt}(2011)}]{Brini2011}%
  \BibitemOpen
  \bibfield  {author} {\bibinfo {author} {\bibfnamefont {E.}~\bibnamefont
  {Brini}}, \bibinfo {author} {\bibfnamefont {V.}~\bibnamefont {Marcon}}, \
  and\ \bibinfo {author} {\bibfnamefont {N.~F.~A.}\ \bibnamefont {van~der
  Vegt}},\ }\href {\doibase 10.1039/C0CP02888F} {\bibfield  {journal} {\bibinfo
   {journal} {Phys. Chem. Chem. Phys.}\ }\textbf {\bibinfo {volume} {13}},\
  \bibinfo {pages} {10468} (\bibinfo {year} {2011})}\BibitemShut {NoStop}%
\bibitem [{\citenamefont {Lyubartsev}\ and\ \citenamefont
  {Laaksonen}(1995)}]{Lyubartsev1995}%
  \BibitemOpen
  \bibfield  {author} {\bibinfo {author} {\bibfnamefont {A.~P.}\ \bibnamefont
  {Lyubartsev}}\ and\ \bibinfo {author} {\bibfnamefont {A.}~\bibnamefont
  {Laaksonen}},\ }\href {\doibase 10.1103/PhysRevE.52.3730} {\bibfield
  {journal} {\bibinfo  {journal} {Phys. Rev. E}\ }\textbf {\bibinfo {volume}
  {52}},\ \bibinfo {pages} {3730} (\bibinfo {year} {1995})}\BibitemShut
  {NoStop}%
\bibitem [{\citenamefont {Lyubartsev}\ and\ \citenamefont
  {Laaksonen}(1997)}]{Lyubartsev1997}%
  \BibitemOpen
  \bibfield  {author} {\bibinfo {author} {\bibfnamefont {A.~P.}\ \bibnamefont
  {Lyubartsev}}\ and\ \bibinfo {author} {\bibfnamefont {A.}~\bibnamefont
  {Laaksonen}},\ }\href {\doibase 10.1103/PhysRevE.55.5689} {\bibfield
  {journal} {\bibinfo  {journal} {Phys. Rev. E}\ }\textbf {\bibinfo {volume}
  {55}},\ \bibinfo {pages} {5689} (\bibinfo {year} {1997})}\BibitemShut
  {NoStop}%
\bibitem [{\citenamefont {Soper}(1996)}]{Soper1996}%
  \BibitemOpen
  \bibfield  {author} {\bibinfo {author} {\bibfnamefont {A.}~\bibnamefont
  {Soper}},\ }\href {\doibase http://dx.doi.org/10.1016/0301-0104(95)00357-6}
  {\bibfield  {journal} {\bibinfo  {journal} {Chem. Phys.}\ }\textbf {\bibinfo
  {volume} {202}},\ \bibinfo {pages} {295 } (\bibinfo {year}
  {1996})}\BibitemShut {NoStop}%
\bibitem [{\citenamefont {M\"uller-Plathe}(2002)}]{MullerPlathe2002}%
  \BibitemOpen
  \bibfield  {author} {\bibinfo {author} {\bibfnamefont {F.}~\bibnamefont
  {M\"uller-Plathe}},\ }\href {\doibase
  10.1002/1439-7641(20020916)3:9<754::AID-CPHC754>3.0.CO;2-U} {\bibfield
  {journal} {\bibinfo  {journal} {ChemPhysChem}\ }\textbf {\bibinfo {volume}
  {3}},\ \bibinfo {pages} {754} (\bibinfo {year} {2002})}\BibitemShut {NoStop}%
\bibitem [{\citenamefont {R\"uhle}\ and\ \citenamefont
  {Junghans}(2011)}]{Ruhle2011}%
  \BibitemOpen
  \bibfield  {author} {\bibinfo {author} {\bibfnamefont {V.}~\bibnamefont
  {R\"uhle}}\ and\ \bibinfo {author} {\bibfnamefont {C.}~\bibnamefont
  {Junghans}},\ }\href {\doibase 10.1002/mats.201100011} {\bibfield  {journal}
  {\bibinfo  {journal} {Macromol. Theory Simul.}\ }\textbf {\bibinfo {volume}
  {20}},\ \bibinfo {pages} {472} (\bibinfo {year} {2011})}\BibitemShut
  {NoStop}%
\bibitem [{\citenamefont {Izvekov}\ and\ \citenamefont
  {Voth}(2006)}]{Izvekov2006}%
  \BibitemOpen
  \bibfield  {author} {\bibinfo {author} {\bibfnamefont {S.}~\bibnamefont
  {Izvekov}}\ and\ \bibinfo {author} {\bibfnamefont {G.~A.}\ \bibnamefont
  {Voth}},\ }\href {\doibase 10.1063/1.2360580} {\bibfield  {journal} {\bibinfo
   {journal} {J. Chem. Phys.}\ }\textbf {\bibinfo {volume} {125}},\ \bibinfo
  {pages} {151101} (\bibinfo {year} {2006})}\BibitemShut {NoStop}%
\bibitem [{\citenamefont {Davtyan}\ \emph {et~al.}(2015)\citenamefont
  {Davtyan}, \citenamefont {Dama}, \citenamefont {Voth},\ and\ \citenamefont
  {Andersen}}]{Davtyan2015}%
  \BibitemOpen
  \bibfield  {author} {\bibinfo {author} {\bibfnamefont {A.}~\bibnamefont
  {Davtyan}}, \bibinfo {author} {\bibfnamefont {J.~F.}\ \bibnamefont {Dama}},
  \bibinfo {author} {\bibfnamefont {G.~A.}\ \bibnamefont {Voth}}, \ and\
  \bibinfo {author} {\bibfnamefont {H.~C.}\ \bibnamefont {Andersen}},\
  }\href@noop {} {\bibfield  {journal} {\bibinfo  {journal} {J. Chem. Phys.}\
  }\textbf {\bibinfo {volume} {142}},\ \bibinfo {pages} {154104} (\bibinfo
  {year} {2015})}\BibitemShut {NoStop}%
\bibitem [{\citenamefont {Nielsen}\ \emph {et~al.}(2003)\citenamefont
  {Nielsen}, \citenamefont {Lopez}, \citenamefont {Srinivas},\ and\
  \citenamefont {Klein}}]{Nielsen2003}%
  \BibitemOpen
  \bibfield  {author} {\bibinfo {author} {\bibfnamefont {S.~O.}\ \bibnamefont
  {Nielsen}}, \bibinfo {author} {\bibfnamefont {C.~F.}\ \bibnamefont {Lopez}},
  \bibinfo {author} {\bibfnamefont {G.}~\bibnamefont {Srinivas}}, \ and\
  \bibinfo {author} {\bibfnamefont {M.~L.}\ \bibnamefont {Klein}},\ }\href
  {\doibase 10.1063/1.1607955} {\bibfield  {journal} {\bibinfo  {journal} {J.
  Chem. Phys.}\ }\textbf {\bibinfo {volume} {119}},\ \bibinfo {pages} {7043}
  (\bibinfo {year} {2003})}\BibitemShut {NoStop}%
\bibitem [{\citenamefont {Bernabei}\ \emph {et~al.}(2013)\citenamefont
  {Bernabei}, \citenamefont {Bacova}, \citenamefont {Moreno}, \citenamefont
  {Narros},\ and\ \citenamefont {Likos}}]{Bernabei2013}%
  \BibitemOpen
  \bibfield  {author} {\bibinfo {author} {\bibfnamefont {M.}~\bibnamefont
  {Bernabei}}, \bibinfo {author} {\bibfnamefont {P.}~\bibnamefont {Bacova}},
  \bibinfo {author} {\bibfnamefont {A.~J.}\ \bibnamefont {Moreno}}, \bibinfo
  {author} {\bibfnamefont {A.}~\bibnamefont {Narros}}, \ and\ \bibinfo {author}
  {\bibfnamefont {C.~N.}\ \bibnamefont {Likos}},\ }\href {\doibase
  10.1039/C2SM27199K} {\bibfield  {journal} {\bibinfo  {journal} {Soft Matter}\
  }\textbf {\bibinfo {volume} {9}},\ \bibinfo {pages} {1287} (\bibinfo {year}
  {2013})}\BibitemShut {NoStop}%
\bibitem [{\citenamefont {Poier}, \citenamefont {Likos},\ and\ \citenamefont
  {Matthews}(2014)}]{Poier2014}%
  \BibitemOpen
  \bibfield  {author} {\bibinfo {author} {\bibfnamefont {P.}~\bibnamefont
  {Poier}}, \bibinfo {author} {\bibfnamefont {C.~N.}\ \bibnamefont {Likos}}, \
  and\ \bibinfo {author} {\bibfnamefont {R.}~\bibnamefont {Matthews}},\ }\href
  {\doibase 10.1021/ma5006414} {\bibfield  {journal} {\bibinfo  {journal}
  {Macromolecules}\ }\textbf {\bibinfo {volume} {47}},\ \bibinfo {pages} {3394}
  (\bibinfo {year} {2014})}\BibitemShut {NoStop}%
\bibitem [{\citenamefont {Poier}\ \emph {et~al.}(2015)\citenamefont {Poier},
  \citenamefont {Likos}, \citenamefont {Moreno},\ and\ \citenamefont
  {Blaak}}]{Poier2015}%
  \BibitemOpen
  \bibfield  {author} {\bibinfo {author} {\bibfnamefont {P.}~\bibnamefont
  {Poier}}, \bibinfo {author} {\bibfnamefont {C.~N.}\ \bibnamefont {Likos}},
  \bibinfo {author} {\bibfnamefont {A.~J.}\ \bibnamefont {Moreno}}, \ and\
  \bibinfo {author} {\bibfnamefont {R.}~\bibnamefont {Blaak}},\ }\href
  {\doibase 10.1021/acs.macromol.5b00603} {\bibfield  {journal} {\bibinfo
  {journal} {Macromolecules}\ }\textbf {\bibinfo {volume} {48}},\ \bibinfo
  {pages} {4983} (\bibinfo {year} {2015})}\BibitemShut {NoStop}%
\bibitem [{\citenamefont {Jorgensen}, \citenamefont {Madura},\ and\
  \citenamefont {Swenson}(1984)}]{Jorgensen1984}%
  \BibitemOpen
  \bibfield  {author} {\bibinfo {author} {\bibfnamefont {W.~L.}\ \bibnamefont
  {Jorgensen}}, \bibinfo {author} {\bibfnamefont {J.~D.}\ \bibnamefont
  {Madura}}, \ and\ \bibinfo {author} {\bibfnamefont {C.~J.}\ \bibnamefont
  {Swenson}},\ }\href {\doibase 10.1021/ja00334a030} {\bibfield  {journal}
  {\bibinfo  {journal} {J. Am. Chem. Soc.}\ }\textbf {\bibinfo {volume}
  {106}},\ \bibinfo {pages} {6638} (\bibinfo {year} {1984})}\BibitemShut
  {NoStop}%
\bibitem [{\citenamefont {von Lilienfeld}\ and\ \citenamefont
  {Andrienko}(2006)}]{Lilienfeld2006}%
  \BibitemOpen
  \bibfield  {author} {\bibinfo {author} {\bibfnamefont {O.~A.}\ \bibnamefont
  {von Lilienfeld}}\ and\ \bibinfo {author} {\bibfnamefont {D.}~\bibnamefont
  {Andrienko}},\ }\href {\doibase 10.1063/1.2162543} {\bibfield  {journal}
  {\bibinfo  {journal} {J. Chem. Phys.}\ }\textbf {\bibinfo {volume} {124}},\
  \bibinfo {pages} {054307} (\bibinfo {year} {2006})}\BibitemShut {NoStop}%
\bibitem [{\citenamefont {Norberg}\ and\ \citenamefont
  {Nilsson}(1995)}]{Norberg1995}%
  \BibitemOpen
  \bibfield  {author} {\bibinfo {author} {\bibfnamefont {J.}~\bibnamefont
  {Norberg}}\ and\ \bibinfo {author} {\bibfnamefont {L.}~\bibnamefont
  {Nilsson}},\ }\href {\doibase 10.1021/ja00149a006} {\bibfield  {journal}
  {\bibinfo  {journal} {J. Am. Chem. Soc.}\ }\textbf {\bibinfo {volume}
  {117}},\ \bibinfo {pages} {10832} (\bibinfo {year} {1995})}\BibitemShut
  {NoStop}%
\bibitem [{\citenamefont {Morriss-Andrews}(2009)}]{MorrissAndrews2009}%
  \BibitemOpen
  \bibfield  {author} {\bibinfo {author} {\bibfnamefont {H.~A.}\ \bibnamefont
  {Morriss-Andrews}},\ }\emph {\bibinfo {title} {Coarse-Grained Molecular
  Dynamics Simulations of {DNA} Representing Bases as Ellipsoids}},\ \href
  {http://hdl.handle.net/2429/12543} {Ph.D. thesis},\ \bibinfo  {school}
  {UNIVERSITY OF BRITISH COLUMBIA (Vancouver)} (\bibinfo {year}
  {2009})\BibitemShut {NoStop}%
\bibitem [{\citenamefont {Morriss-Andrews}, \citenamefont {Rottler},\ and\
  \citenamefont {Plotkin}(2010)}]{MorrissAndrews2010}%
  \BibitemOpen
  \bibfield  {author} {\bibinfo {author} {\bibfnamefont {A.}~\bibnamefont
  {Morriss-Andrews}}, \bibinfo {author} {\bibfnamefont {J.}~\bibnamefont
  {Rottler}}, \ and\ \bibinfo {author} {\bibfnamefont {S.~S.}\ \bibnamefont
  {Plotkin}},\ }\href {\doibase 10.1063/1.3269994} {\bibfield  {journal}
  {\bibinfo  {journal} {J. Chem. Phys.}\ }\textbf {\bibinfo {volume} {132}},\
  \bibinfo {pages} {035105} (\bibinfo {year} {2010})}\BibitemShut {NoStop}%
\bibitem [{\citenamefont {Bennun}\ \emph {et~al.}(2009)\citenamefont {Bennun},
  \citenamefont {Hoopes}, \citenamefont {Xing},\ and\ \citenamefont
  {Faller}}]{Bennun2009}%
  \BibitemOpen
  \bibfield  {author} {\bibinfo {author} {\bibfnamefont {S.~V.}\ \bibnamefont
  {Bennun}}, \bibinfo {author} {\bibfnamefont {M.~I.}\ \bibnamefont {Hoopes}},
  \bibinfo {author} {\bibfnamefont {C.}~\bibnamefont {Xing}}, \ and\ \bibinfo
  {author} {\bibfnamefont {R.}~\bibnamefont {Faller}},\ }\href {\doibase
  10.1016/j.chemphyslip.2009.03.003} {\bibfield  {journal} {\bibinfo  {journal}
  {Chem. Phys. Lipids}\ }\textbf {\bibinfo {volume} {159}},\ \bibinfo {pages}
  {59} (\bibinfo {year} {2009})}\BibitemShut {NoStop}%
\bibitem [{\citenamefont {Masunov}\ and\ \citenamefont
  {Lazaridis}(2003)}]{Masunov2003}%
  \BibitemOpen
  \bibfield  {author} {\bibinfo {author} {\bibfnamefont {A.}~\bibnamefont
  {Masunov}}\ and\ \bibinfo {author} {\bibfnamefont {T.}~\bibnamefont
  {Lazaridis}},\ }\href {\doibase 10.1021/ja025521w} {\bibfield  {journal}
  {\bibinfo  {journal} {J. Am. Chem. Soc.}\ }\textbf {\bibinfo {volume}
  {125}},\ \bibinfo {pages} {1722} (\bibinfo {year} {2003})}\BibitemShut
  {NoStop}%
\bibitem [{\citenamefont {Villa}\ \emph {et~al.}(2004)\citenamefont {Villa},
  \citenamefont {Balaeff}, \citenamefont {Mahadevan},\ and\ \citenamefont
  {Schulten}}]{Villa2004}%
  \BibitemOpen
  \bibfield  {author} {\bibinfo {author} {\bibfnamefont {E.}~\bibnamefont
  {Villa}}, \bibinfo {author} {\bibfnamefont {A.}~\bibnamefont {Balaeff}},
  \bibinfo {author} {\bibfnamefont {L.}~\bibnamefont {Mahadevan}}, \ and\
  \bibinfo {author} {\bibfnamefont {K.}~\bibnamefont {Schulten}},\ }\href
  {\doibase 10.1137/040604789} {\bibfield  {journal} {\bibinfo  {journal}
  {Multiscale Model. Sim.}\ }\textbf {\bibinfo {volume} {2}},\ \bibinfo {pages}
  {527} (\bibinfo {year} {2004})}\BibitemShut {NoStop}%
\bibitem [{\citenamefont {Bhattacherjee}, \citenamefont {Krepel},\ and\
  \citenamefont {Levy}(2016)}]{Bhattacherjee2016}%
  \BibitemOpen
  \bibfield  {author} {\bibinfo {author} {\bibfnamefont {A.}~\bibnamefont
  {Bhattacherjee}}, \bibinfo {author} {\bibfnamefont {D.}~\bibnamefont
  {Krepel}}, \ and\ \bibinfo {author} {\bibfnamefont {Y.}~\bibnamefont
  {Levy}},\ }\href {\doibase 10.1002/wcms.1262} {\bibfield  {journal} {\bibinfo
   {journal} {Wiley Interdiscip. Rev. Comput. Mol. Sci.}\ }\textbf {\bibinfo
  {volume} {6}},\ \bibinfo {pages} {515} (\bibinfo {year} {2016})}\BibitemShut
  {NoStop}%
\bibitem [{\citenamefont {Hagan}\ and\ \citenamefont
  {Zandi}(2016)}]{Hagan2016}%
  \BibitemOpen
  \bibfield  {author} {\bibinfo {author} {\bibfnamefont {M.~F.}\ \bibnamefont
  {Hagan}}\ and\ \bibinfo {author} {\bibfnamefont {R.}~\bibnamefont {Zandi}},\
  }\href {\doibase 10.1016/j.coviro.2016.02.012} {\bibfield  {journal}
  {\bibinfo  {journal} {Curr. Opin. Virol.}\ }\textbf {\bibinfo {volume}
  {18}},\ \bibinfo {pages} {36} (\bibinfo {year} {2016})}\BibitemShut {NoStop}%
\bibitem [{\citenamefont {M\"uller}\ \emph {et~al.}(2014)\citenamefont
  {M\"uller}, \citenamefont {Osterman}, \citenamefont {Babi\v{c}},
  \citenamefont {Likos}, \citenamefont {Dobnikar},\ and\ \citenamefont
  {Nikoubashman}}]{Muller2014}%
  \BibitemOpen
  \bibfield  {author} {\bibinfo {author} {\bibfnamefont {K.}~\bibnamefont
  {M\"uller}}, \bibinfo {author} {\bibfnamefont {N.}~\bibnamefont {Osterman}},
  \bibinfo {author} {\bibfnamefont {D.}~\bibnamefont {Babi\v{c}}}, \bibinfo
  {author} {\bibfnamefont {C.~N.}\ \bibnamefont {Likos}}, \bibinfo {author}
  {\bibfnamefont {J.}~\bibnamefont {Dobnikar}}, \ and\ \bibinfo {author}
  {\bibfnamefont {A.}~\bibnamefont {Nikoubashman}},\ }\href {\doibase
  10.1021/la500896e} {\bibfield  {journal} {\bibinfo  {journal} {Langmuir}\
  }\textbf {\bibinfo {volume} {30}},\ \bibinfo {pages} {5088} (\bibinfo {year}
  {2014})}\BibitemShut {NoStop}%
\bibitem [{\citenamefont {Li}\ \emph {et~al.}(2005)\citenamefont {Li},
  \citenamefont {Harnau}, \citenamefont {Rosenfeldt},\ and\ \citenamefont
  {Ballauff}}]{Li2005}%
  \BibitemOpen
  \bibfield  {author} {\bibinfo {author} {\bibfnamefont {L.}~\bibnamefont
  {Li}}, \bibinfo {author} {\bibfnamefont {L.}~\bibnamefont {Harnau}}, \bibinfo
  {author} {\bibfnamefont {S.}~\bibnamefont {Rosenfeldt}}, \ and\ \bibinfo
  {author} {\bibfnamefont {M.}~\bibnamefont {Ballauff}},\ }\href {\doibase
  10.1103/PhysRevE.72.051504} {\bibfield  {journal} {\bibinfo  {journal} {Phys.
  Rev. E}\ }\textbf {\bibinfo {volume} {72}},\ \bibinfo {pages} {051504}
  (\bibinfo {year} {2005})}\BibitemShut {NoStop}%
\bibitem [{\citenamefont {Likos}(2001)}]{Likos2001}%
  \BibitemOpen
  \bibfield  {author} {\bibinfo {author} {\bibfnamefont {C.~N.}\ \bibnamefont
  {Likos}},\ }\href {\doibase 10.1016/S0370-1573(00)00141-1} {\bibfield
  {journal} {\bibinfo  {journal} {Physics Reports}\ }\textbf {\bibinfo {volume}
  {348}},\ \bibinfo {pages} {267} (\bibinfo {year} {2001})}\BibitemShut
  {NoStop}%
\bibitem [{\citenamefont {Sippl}(1990)}]{Sippl1990}%
  \BibitemOpen
  \bibfield  {author} {\bibinfo {author} {\bibfnamefont {M.~J.}\ \bibnamefont
  {Sippl}},\ }\href {\doibase 10.1016/S0022-2836(05)80269-4} {\bibfield
  {journal} {\bibinfo  {journal} {J. Mol. Biol.}\ }\textbf {\bibinfo {volume}
  {213}},\ \bibinfo {pages} {859} (\bibinfo {year} {1990})}\BibitemShut
  {NoStop}%
\bibitem [{\citenamefont {Buchete}, \citenamefont {Straub},\ and\ \citenamefont
  {Thirumalai}(2004)}]{Buchete2004}%
  \BibitemOpen
  \bibfield  {author} {\bibinfo {author} {\bibfnamefont {N.-V.}\ \bibnamefont
  {Buchete}}, \bibinfo {author} {\bibfnamefont {J.~E.}\ \bibnamefont {Straub}},
  \ and\ \bibinfo {author} {\bibfnamefont {D.}~\bibnamefont {Thirumalai}},\
  }\href {\doibase 10.1110/ps.03488704} {\bibfield  {journal} {\bibinfo
  {journal} {Protein Sci.}\ }\textbf {\bibinfo {volume} {13}},\ \bibinfo
  {pages} {862} (\bibinfo {year} {2004})}\BibitemShut {NoStop}%
\bibitem [{\citenamefont {Zhou}\ and\ \citenamefont
  {Skolnick}(2011)}]{Zhou2011}%
  \BibitemOpen
  \bibfield  {author} {\bibinfo {author} {\bibfnamefont {H.}~\bibnamefont
  {Zhou}}\ and\ \bibinfo {author} {\bibfnamefont {J.}~\bibnamefont
  {Skolnick}},\ }\href {\doibase 10.1016/j.bpj.2011.09.012} {\bibfield
  {journal} {\bibinfo  {journal} {Biophys. J.}\ }\textbf {\bibinfo {volume}
  {101}},\ \bibinfo {pages} {2043} (\bibinfo {year} {2011})}\BibitemShut
  {NoStop}%
\bibitem [{\citenamefont {Heinemann}\ \emph {et~al.}(2014)\citenamefont
  {Heinemann}, \citenamefont {Palczynski}, \citenamefont {Dzubiella},\ and\
  \citenamefont {Klapp}}]{Heinemann2014}%
  \BibitemOpen
  \bibfield  {author} {\bibinfo {author} {\bibfnamefont {T.}~\bibnamefont
  {Heinemann}}, \bibinfo {author} {\bibfnamefont {K.}~\bibnamefont
  {Palczynski}}, \bibinfo {author} {\bibfnamefont {J.}~\bibnamefont
  {Dzubiella}}, \ and\ \bibinfo {author} {\bibfnamefont {S.~H.~L.}\
  \bibnamefont {Klapp}},\ }\href {\doibase 10.1063/1.4902824} {\bibfield
  {journal} {\bibinfo  {journal} {J. Chem. Phys.}\ }\textbf {\bibinfo {volume}
  {141}},\ \bibinfo {pages} {214110} (\bibinfo {year} {2014})}\BibitemShut
  {NoStop}%
\bibitem [{\citenamefont {Heinemann}(2016)}]{Heinemannphd2016}%
  \BibitemOpen
  \bibfield  {author} {\bibinfo {author} {\bibfnamefont {T.}~\bibnamefont
  {Heinemann}},\ }\emph {\bibinfo {title} {Systematic coarse-graining
  procedures for molecular systems}},\ \href {\doibase
  10.14279/depositonce-5058} {Ph.D. thesis},\ \bibinfo  {school} {Technical
  University Berlin} (\bibinfo {year} {2016})\BibitemShut {NoStop}%
\bibitem [{\citenamefont {Tsch\"op}\ \emph
  {et~al.}(1998{\natexlab{a}})\citenamefont {Tsch\"op}, \citenamefont {Kremer},
  \citenamefont {Batoulis}, \citenamefont {B\"urger},\ and\ \citenamefont
  {Hahn}}]{Tschoep1998}%
  \BibitemOpen
  \bibfield  {author} {\bibinfo {author} {\bibfnamefont {W.}~\bibnamefont
  {Tsch\"op}}, \bibinfo {author} {\bibfnamefont {K.}~\bibnamefont {Kremer}},
  \bibinfo {author} {\bibfnamefont {J.}~\bibnamefont {Batoulis}}, \bibinfo
  {author} {\bibfnamefont {T.}~\bibnamefont {B\"urger}}, \ and\ \bibinfo
  {author} {\bibfnamefont {O.}~\bibnamefont {Hahn}},\ }\href {\doibase
  10.1002/(SICI)1521-4044(199802)49:2/3<61::AID-APOL61>3.0.CO;2-V} {\bibfield
  {journal} {\bibinfo  {journal} {Acta Polym.}\ }\textbf {\bibinfo {volume}
  {49}},\ \bibinfo {pages} {61} (\bibinfo {year}
  {1998}{\natexlab{a}})}\BibitemShut {NoStop}%
\bibitem [{\citenamefont {Tsch\"op}\ \emph
  {et~al.}(1998{\natexlab{b}})\citenamefont {Tsch\"op}, \citenamefont {Kremer},
  \citenamefont {Hahn}, \citenamefont {Batoulis},\ and\ \citenamefont
  {B\"urger}}]{Tschoep1998b}%
  \BibitemOpen
  \bibfield  {author} {\bibinfo {author} {\bibfnamefont {W.}~\bibnamefont
  {Tsch\"op}}, \bibinfo {author} {\bibfnamefont {K.}~\bibnamefont {Kremer}},
  \bibinfo {author} {\bibfnamefont {O.}~\bibnamefont {Hahn}}, \bibinfo {author}
  {\bibfnamefont {J.}~\bibnamefont {Batoulis}}, \ and\ \bibinfo {author}
  {\bibfnamefont {T.}~\bibnamefont {B\"urger}},\ }\href {\doibase
  10.1002/(SICI)1521-4044(199802)49:2/3<75::AID-APOL75>3.0.CO;2-5} {\bibfield
  {journal} {\bibinfo  {journal} {Acta Polym.}\ }\textbf {\bibinfo {volume}
  {49}},\ \bibinfo {pages} {75} (\bibinfo {year}
  {1998}{\natexlab{b}})}\BibitemShut {NoStop}%
\bibitem [{\citenamefont {Heinemann}\ \emph {et~al.}(2015)\citenamefont
  {Heinemann}, \citenamefont {Palczynski}, \citenamefont {Dzubiella},\ and\
  \citenamefont {Klapp}}]{Heinemann2015}%
  \BibitemOpen
  \bibfield  {author} {\bibinfo {author} {\bibfnamefont {T.}~\bibnamefont
  {Heinemann}}, \bibinfo {author} {\bibfnamefont {K.}~\bibnamefont
  {Palczynski}}, \bibinfo {author} {\bibfnamefont {J.}~\bibnamefont
  {Dzubiella}}, \ and\ \bibinfo {author} {\bibfnamefont {S.~H.~L.}\
  \bibnamefont {Klapp}},\ }\href {\doibase 10.1063/1.4935063} {\bibfield
  {journal} {\bibinfo  {journal} {J. Chem. Phys.}\ }\textbf {\bibinfo {volume}
  {143}},\ \bibinfo {pages} {174110} (\bibinfo {year} {2015})}\BibitemShut
  {NoStop}%
\bibitem [{\citenamefont {Hernandez-Rojas}, \citenamefont {Calvo},\ and\
  \citenamefont {Wales}(2016)}]{HernandezRojas2016}%
  \BibitemOpen
  \bibfield  {author} {\bibinfo {author} {\bibfnamefont {J.}~\bibnamefont
  {Hernandez-Rojas}}, \bibinfo {author} {\bibfnamefont {F.}~\bibnamefont
  {Calvo}}, \ and\ \bibinfo {author} {\bibfnamefont {D.~J.}\ \bibnamefont
  {Wales}},\ }\href {\doibase 10.1039/C6CP00592F} {\bibfield  {journal}
  {\bibinfo  {journal} {Phys. Chem. Chem. Phys.}\ }\textbf {\bibinfo {volume}
  {18}},\ \bibinfo {pages} {13736} (\bibinfo {year} {2016})}\BibitemShut
  {NoStop}%
\bibitem [{\citenamefont {Zacharopoulos}, \citenamefont {Vergadou},\ and\
  \citenamefont {Theodorou}(2005)}]{Zacharopoulos2005}%
  \BibitemOpen
  \bibfield  {author} {\bibinfo {author} {\bibfnamefont {N.}~\bibnamefont
  {Zacharopoulos}}, \bibinfo {author} {\bibfnamefont {N.}~\bibnamefont
  {Vergadou}}, \ and\ \bibinfo {author} {\bibfnamefont {D.~N.}\ \bibnamefont
  {Theodorou}},\ }\href@noop {} {\bibfield  {journal} {\bibinfo  {journal} {J.
  Chem. Phys.}\ }\textbf {\bibinfo {volume} {122}},\ \bibinfo {pages} {244111}
  (\bibinfo {year} {2005})}\BibitemShut {NoStop}%
\bibitem [{\citenamefont {Lettieri}\ and\ \citenamefont
  {Zuckerman}(2012)}]{Lettieri2012}%
  \BibitemOpen
  \bibfield  {author} {\bibinfo {author} {\bibfnamefont {S.}~\bibnamefont
  {Lettieri}}\ and\ \bibinfo {author} {\bibfnamefont {D.~M.}\ \bibnamefont
  {Zuckerman}},\ }\href {\doibase 10.1002/jcc.21970} {\bibfield  {journal}
  {\bibinfo  {journal} {J. Comput. Chem.}\ }\textbf {\bibinfo {volume} {33}},\
  \bibinfo {pages} {268} (\bibinfo {year} {2012})}\BibitemShut {NoStop}%
\bibitem [{\citenamefont {Spiriti}\ and\ \citenamefont
  {Zuckerman}(2014)}]{Spiriti2014}%
  \BibitemOpen
  \bibfield  {author} {\bibinfo {author} {\bibfnamefont {J.}~\bibnamefont
  {Spiriti}}\ and\ \bibinfo {author} {\bibfnamefont {D.~M.}\ \bibnamefont
  {Zuckerman}},\ }\href {\doibase 10.1021/ct500622z} {\bibfield  {journal}
  {\bibinfo  {journal} {J. Chem. Theory Comput.}\ }\textbf {\bibinfo {volume}
  {10}},\ \bibinfo {pages} {5161} (\bibinfo {year} {2014})}\BibitemShut
  {NoStop}%
\bibitem [{\citenamefont {Spiriti}\ and\ \citenamefont
  {Zuckerman}(2015)}]{Spiriti2015}%
  \BibitemOpen
  \bibfield  {author} {\bibinfo {author} {\bibfnamefont {J.}~\bibnamefont
  {Spiriti}}\ and\ \bibinfo {author} {\bibfnamefont {D.~M.}\ \bibnamefont
  {Zuckerman}},\ }\href@noop {} {\bibfield  {journal} {\bibinfo  {journal} {J.
  Chem. Phys.}\ }\textbf {\bibinfo {volume} {143}},\ \bibinfo {eid} {243159}
  (\bibinfo {year} {2015})}\BibitemShut {NoStop}%
\bibitem [{\citenamefont {Louis}(2002)}]{Louis2002}%
  \BibitemOpen
  \bibfield  {author} {\bibinfo {author} {\bibfnamefont {A.~A.}\ \bibnamefont
  {Louis}},\ }\href@noop {} {\bibfield  {journal} {\bibinfo  {journal} {J.
  Phys. Condens. Matter}\ }\textbf {\bibinfo {volume} {14}},\ \bibinfo {pages}
  {9187} (\bibinfo {year} {2002})}\BibitemShut {NoStop}%
\bibitem [{\citenamefont {Axilrod}\ and\ \citenamefont
  {Teller}(1943)}]{Axilrod1943}%
  \BibitemOpen
  \bibfield  {author} {\bibinfo {author} {\bibfnamefont {B.}~\bibnamefont
  {Axilrod}}\ and\ \bibinfo {author} {\bibfnamefont {E.}~\bibnamefont
  {Teller}},\ }\href@noop {} {\bibfield  {journal} {\bibinfo  {journal} {J.
  Chem. Phys.}\ }\textbf {\bibinfo {volume} {11}},\ \bibinfo {pages} {299}
  (\bibinfo {year} {1943})}\BibitemShut {NoStop}%
\bibitem [{\citenamefont {Evans}\ and\ \citenamefont
  {Watts}(1976)}]{Evans1976}%
  \BibitemOpen
  \bibfield  {author} {\bibinfo {author} {\bibfnamefont {D.}~\bibnamefont
  {Evans}}\ and\ \bibinfo {author} {\bibfnamefont {R.}~\bibnamefont {Watts}},\
  }\href {\doibase 10.1080/00268977600100071} {\bibfield  {journal} {\bibinfo
  {journal} {Mol. Phys.}\ }\textbf {\bibinfo {volume} {31}},\ \bibinfo {pages}
  {83} (\bibinfo {year} {1976})}\BibitemShut {NoStop}%
\bibitem [{\citenamefont {Kalligiannaki}\ \emph {et~al.}(2016)\citenamefont
  {Kalligiannaki}, \citenamefont {Chazirakis}, \citenamefont {Tsourtis},
  \citenamefont {Katsoulakis}, \citenamefont {Plech{\'a}{\v{c}}},\ and\
  \citenamefont {Harmandaris}}]{Kalligiannaki2016}%
  \BibitemOpen
  \bibfield  {author} {\bibinfo {author} {\bibfnamefont {E.}~\bibnamefont
  {Kalligiannaki}}, \bibinfo {author} {\bibfnamefont {A.}~\bibnamefont
  {Chazirakis}}, \bibinfo {author} {\bibfnamefont {A.}~\bibnamefont
  {Tsourtis}}, \bibinfo {author} {\bibfnamefont {M.}~\bibnamefont
  {Katsoulakis}}, \bibinfo {author} {\bibfnamefont {P.}~\bibnamefont
  {Plech{\'a}{\v{c}}}}, \ and\ \bibinfo {author} {\bibfnamefont
  {V.}~\bibnamefont {Harmandaris}},\ }\href@noop {} {\bibfield  {journal}
  {\bibinfo  {journal} {Eur. Phys. J.}\ }\textbf {\bibinfo {volume} {225}},\
  \bibinfo {pages} {1347} (\bibinfo {year} {2016})}\BibitemShut {NoStop}%
\bibitem [{\citenamefont {Kirkwood}(1935)}]{Kirkwood1935}%
  \BibitemOpen
  \bibfield  {author} {\bibinfo {author} {\bibfnamefont {J.~G.}\ \bibnamefont
  {Kirkwood}},\ }\href {\doibase 10.1063/1.1749657} {\bibfield  {journal}
  {\bibinfo  {journal} {J. Chem. Phys.}\ }\textbf {\bibinfo {volume} {3}},\
  \bibinfo {pages} {300} (\bibinfo {year} {1935})}\BibitemShut {NoStop}%
\bibitem [{\citenamefont {Fan}\ \emph {et~al.}(2011)\citenamefont {Fan},
  \citenamefont {Schneidman-Duhovny}, \citenamefont {Irwin}, \citenamefont
  {Dong}, \citenamefont {Shoichet},\ and\ \citenamefont {Sali}}]{Fan2011}%
  \BibitemOpen
  \bibfield  {author} {\bibinfo {author} {\bibfnamefont {H.}~\bibnamefont
  {Fan}}, \bibinfo {author} {\bibfnamefont {D.}~\bibnamefont
  {Schneidman-Duhovny}}, \bibinfo {author} {\bibfnamefont {J.~J.}\ \bibnamefont
  {Irwin}}, \bibinfo {author} {\bibfnamefont {G.}~\bibnamefont {Dong}},
  \bibinfo {author} {\bibfnamefont {B.~K.}\ \bibnamefont {Shoichet}}, \ and\
  \bibinfo {author} {\bibfnamefont {A.}~\bibnamefont {Sali}},\ }\href {\doibase
  10.1021/ci200377u} {\bibfield  {journal} {\bibinfo  {journal} {J. Chem. Inf.
  Model.}\ }\textbf {\bibinfo {volume} {51}},\ \bibinfo {pages} {3078}
  (\bibinfo {year} {2011})}\BibitemShut {NoStop}%
\bibitem [{\citenamefont {Philips}\ \emph {et~al.}(2013)\citenamefont
  {Philips}, \citenamefont {Milanowska}, \citenamefont {Lach},\ and\
  \citenamefont {Bujnicki}}]{Philips2013}%
  \BibitemOpen
  \bibfield  {author} {\bibinfo {author} {\bibfnamefont {A.}~\bibnamefont
  {Philips}}, \bibinfo {author} {\bibfnamefont {K.}~\bibnamefont {Milanowska}},
  \bibinfo {author} {\bibfnamefont {G.}~\bibnamefont {Lach}}, \ and\ \bibinfo
  {author} {\bibfnamefont {J.}~\bibnamefont {Bujnicki}},\ }\href@noop {}
  {\bibfield  {journal} {\bibinfo  {journal} {RNA (New York, NY)}\ }\textbf
  {\bibinfo {volume} {19}},\ \bibinfo {pages} {1605} (\bibinfo {year}
  {2013})}\BibitemShut {NoStop}%
\bibitem [{\citenamefont {Liu}(2015)}]{Liu2015}%
  \BibitemOpen
  \bibfield  {author} {\bibinfo {author} {\bibfnamefont {H.}~\bibnamefont
  {Liu}},\ }\href {\doibase 10.1007/s40484-015-0054-x} {\bibfield  {journal}
  {\bibinfo  {journal} {Quant. Biol.}\ }\textbf {\bibinfo {volume} {3}},\
  \bibinfo {pages} {157} (\bibinfo {year} {2015})}\BibitemShut {NoStop}%
\bibitem [{\citenamefont {Babadi}, \citenamefont {Everaers},\ and\
  \citenamefont {Ejtehadi}(2006)}]{Babadi2006}%
  \BibitemOpen
  \bibfield  {author} {\bibinfo {author} {\bibfnamefont {M.}~\bibnamefont
  {Babadi}}, \bibinfo {author} {\bibfnamefont {R.}~\bibnamefont {Everaers}}, \
  and\ \bibinfo {author} {\bibfnamefont {M.~R.}\ \bibnamefont {Ejtehadi}},\
  }\href {\doibase 10.1063/1.2179075} {\bibfield  {journal} {\bibinfo
  {journal} {J. Chem. Phys.}\ }\textbf {\bibinfo {volume} {124}},\ \bibinfo
  {pages} {174708} (\bibinfo {year} {2006})}\BibitemShut {NoStop}%
\bibitem [{\citenamefont {Hess}\ and\ \citenamefont {Scheek}(2003)}]{Hess2003}%
  \BibitemOpen
  \bibfield  {author} {\bibinfo {author} {\bibfnamefont {B.}~\bibnamefont
  {Hess}}\ and\ \bibinfo {author} {\bibfnamefont {R.}~\bibnamefont {Scheek}},\
  }\href {\doibase 10.1016/S1090-7807(03)00178-2} {\bibfield  {journal}
  {\bibinfo  {journal} {J. Magn. Reson.}\ }\textbf {\bibinfo {volume} {164}},\
  \bibinfo {pages} {19} (\bibinfo {year} {2003})}\BibitemShut {NoStop}%
\bibitem [{\citenamefont {Car}\ and\ \citenamefont
  {Parrinello}(1985)}]{Car1985}%
  \BibitemOpen
  \bibfield  {author} {\bibinfo {author} {\bibfnamefont {R.}~\bibnamefont
  {Car}}\ and\ \bibinfo {author} {\bibfnamefont {M.}~\bibnamefont
  {Parrinello}},\ }\href {\doibase 10.1103/PhysRevLett.55.2471} {\bibfield
  {journal} {\bibinfo  {journal} {Phys. Rev. Lett.}\ }\textbf {\bibinfo
  {volume} {55}},\ \bibinfo {pages} {2471} (\bibinfo {year}
  {1985})}\BibitemShut {NoStop}%
\bibitem [{\citenamefont {Marx}\ and\ \citenamefont {Hutter}(2000)}]{Marx2000}%
  \BibitemOpen
  \bibfield  {author} {\bibinfo {author} {\bibfnamefont {D.}~\bibnamefont
  {Marx}}\ and\ \bibinfo {author} {\bibfnamefont {J.}~\bibnamefont {Hutter}},\
  }\href@noop {} {\bibfield  {journal} {\bibinfo  {journal} {Modern Methods and
  Algorithms of Quantum Chemistry (FZ Jülich: John von Neumann Institute for
  Computing)}\ }\textbf {\bibinfo {volume} {1}},\ \bibinfo {pages} {301}
  (\bibinfo {year} {2000})}\BibitemShut {NoStop}%
\bibitem [{\citenamefont {Della~Sala}, \citenamefont {Blumstengel},\ and\
  \citenamefont {Henneberger}(2011)}]{DellaSala2011}%
  \BibitemOpen
  \bibfield  {author} {\bibinfo {author} {\bibfnamefont {F.}~\bibnamefont
  {Della~Sala}}, \bibinfo {author} {\bibfnamefont {S.}~\bibnamefont
  {Blumstengel}}, \ and\ \bibinfo {author} {\bibfnamefont {F.}~\bibnamefont
  {Henneberger}},\ }\href {\doibase 10.1103/PhysRevLett.107.146401} {\bibfield
  {journal} {\bibinfo  {journal} {Phys. Rev. Lett.}\ }\textbf {\bibinfo
  {volume} {107}},\ \bibinfo {pages} {146401} (\bibinfo {year}
  {2011})}\BibitemShut {NoStop}%
\bibitem [{\citenamefont {Palczynski}\ and\ \citenamefont
  {Dzubiella}(2014)}]{Palczynski2014b}%
  \BibitemOpen
  \bibfield  {author} {\bibinfo {author} {\bibfnamefont {K.}~\bibnamefont
  {Palczynski}}\ and\ \bibinfo {author} {\bibfnamefont {J.}~\bibnamefont
  {Dzubiella}},\ }\href {\doibase 10.1021/jp507776h} {\bibfield  {journal}
  {\bibinfo  {journal} {J. Phys. Chem. C}\ }\textbf {\bibinfo {volume} {118}},\
  \bibinfo {pages} {26368} (\bibinfo {year} {2014})}\BibitemShut {NoStop}%
\bibitem [{\citenamefont {Kleppmann}\ and\ \citenamefont
  {Klapp}(2015)}]{Kleppmann2015}%
  \BibitemOpen
  \bibfield  {author} {\bibinfo {author} {\bibfnamefont {N.}~\bibnamefont
  {Kleppmann}}\ and\ \bibinfo {author} {\bibfnamefont {S.~H.~L.}\ \bibnamefont
  {Klapp}},\ }\href {\doibase 10.1063/1.4907037} {\bibfield  {journal}
  {\bibinfo  {journal} {J. Chem. Phys.}\ }\textbf {\bibinfo {volume} {142}},\
  \bibinfo {pages} {064701} (\bibinfo {year} {2015})}\BibitemShut {NoStop}%
\bibitem [{\citenamefont {Straatsma}, \citenamefont {Zacharias},\ and\
  \citenamefont {McCammon}(1992)}]{Straatsma1992}%
  \BibitemOpen
  \bibfield  {author} {\bibinfo {author} {\bibfnamefont {T.}~\bibnamefont
  {Straatsma}}, \bibinfo {author} {\bibfnamefont {M.}~\bibnamefont
  {Zacharias}}, \ and\ \bibinfo {author} {\bibfnamefont {J.}~\bibnamefont
  {McCammon}},\ }\href {\doibase 10.1016/0009-2614(92)85971-C} {\bibfield
  {journal} {\bibinfo  {journal} {Chem. Phys. Lett.}\ }\textbf {\bibinfo
  {volume} {196}},\ \bibinfo {pages} {297} (\bibinfo {year}
  {1992})}\BibitemShut {NoStop}%
\bibitem [{\citenamefont {Torrie}\ and\ \citenamefont
  {Valleau}(1974)}]{Torrie1974}%
  \BibitemOpen
  \bibfield  {author} {\bibinfo {author} {\bibfnamefont {G.~M.}\ \bibnamefont
  {Torrie}}\ and\ \bibinfo {author} {\bibfnamefont {J.~P.}\ \bibnamefont
  {Valleau}},\ }\href {\doibase 10.1016/0009-2614(74)80109-0} {\bibfield
  {journal} {\bibinfo  {journal} {Chem. Phys. Lett.}\ }\textbf {\bibinfo
  {volume} {28}},\ \bibinfo {pages} {578} (\bibinfo {year} {1974})}\BibitemShut
  {NoStop}%
\bibitem [{\citenamefont {Torrie}\ and\ \citenamefont
  {Valleau}(1977)}]{Torrie1977}%
  \BibitemOpen
  \bibfield  {author} {\bibinfo {author} {\bibfnamefont {G.~M.}\ \bibnamefont
  {Torrie}}\ and\ \bibinfo {author} {\bibfnamefont {J.~P.}\ \bibnamefont
  {Valleau}},\ }\href {\doibase 10.1016/0021-9991(77)90121-8} {\bibfield
  {journal} {\bibinfo  {journal} {J. Comput. Phys.}\ }\textbf {\bibinfo
  {volume} {23}},\ \bibinfo {pages} {187} (\bibinfo {year} {1977})}\BibitemShut
  {NoStop}%
\bibitem [{\citenamefont {Laio}\ and\ \citenamefont
  {Parrinello}(2002)}]{Laio2002}%
  \BibitemOpen
  \bibfield  {author} {\bibinfo {author} {\bibfnamefont {A.}~\bibnamefont
  {Laio}}\ and\ \bibinfo {author} {\bibfnamefont {M.}~\bibnamefont
  {Parrinello}},\ }\href@noop {} {\bibfield  {journal} {\bibinfo  {journal}
  {Proc. Natl. Acad. Sci. USA}\ }\textbf {\bibinfo {volume} {99}},\ \bibinfo
  {pages} {12562} (\bibinfo {year} {2002})}\BibitemShut {NoStop}%
\bibitem [{\citenamefont {Chodera}\ \emph {et~al.}(2007)\citenamefont
  {Chodera}, \citenamefont {Swope}, \citenamefont {Pitera}, \citenamefont
  {Seok},\ and\ \citenamefont {Dill}}]{Chodera2007}%
  \BibitemOpen
  \bibfield  {author} {\bibinfo {author} {\bibfnamefont {J.~D.}\ \bibnamefont
  {Chodera}}, \bibinfo {author} {\bibfnamefont {W.~C.}\ \bibnamefont {Swope}},
  \bibinfo {author} {\bibfnamefont {J.~W.}\ \bibnamefont {Pitera}}, \bibinfo
  {author} {\bibfnamefont {C.}~\bibnamefont {Seok}}, \ and\ \bibinfo {author}
  {\bibfnamefont {K.~A.}\ \bibnamefont {Dill}},\ }\href {\doibase
  10.1021/ct0502864} {\bibfield  {journal} {\bibinfo  {journal} {J. Chem.
  Theory Comput.}\ }\textbf {\bibinfo {volume} {3}},\ \bibinfo {pages} {26}
  (\bibinfo {year} {2007})}\BibitemShut {NoStop}%
\bibitem [{\citenamefont {Mezei}(1987)}]{Mezei1987}%
  \BibitemOpen
  \bibfield  {author} {\bibinfo {author} {\bibfnamefont {M.}~\bibnamefont
  {Mezei}},\ }\href {\doibase 10.1016/0021-9991(87)90054-4} {\bibfield
  {journal} {\bibinfo  {journal} {J. Comput. Phys.}\ }\textbf {\bibinfo
  {volume} {68}},\ \bibinfo {pages} {237} (\bibinfo {year} {1987})}\BibitemShut
  {NoStop}%
\bibitem [{\citenamefont {K{\"a}stner}\ and\ \citenamefont
  {Thiel}(2005)}]{Kastner2005}%
  \BibitemOpen
  \bibfield  {author} {\bibinfo {author} {\bibfnamefont {J.}~\bibnamefont
  {K{\"a}stner}}\ and\ \bibinfo {author} {\bibfnamefont {W.}~\bibnamefont
  {Thiel}},\ }\href {\doibase 10.1063/1.2052648} {\bibfield  {journal}
  {\bibinfo  {journal} {J. Chem. Phys.}\ }\textbf {\bibinfo {volume} {123}},\
  \bibinfo {pages} {144104} (\bibinfo {year} {2005})}\BibitemShut {NoStop}%
\bibitem [{\citenamefont {K{\"a}stner}(2009)}]{Kastner2009}%
  \BibitemOpen
  \bibfield  {author} {\bibinfo {author} {\bibfnamefont {J.}~\bibnamefont
  {K{\"a}stner}},\ }\href {\doibase 10.1063/1.3175798} {\bibfield  {journal}
  {\bibinfo  {journal} {J. Chem. Phys.}\ }\textbf {\bibinfo {volume} {131}},\
  \bibinfo {pages} {034109} (\bibinfo {year} {2009})}\BibitemShut {NoStop}%
\bibitem [{\citenamefont {Park}\ \emph {et~al.}(2003)\citenamefont {Park},
  \citenamefont {Khalili-Araghi}, \citenamefont {Tajkhorshid},\ and\
  \citenamefont {Schulten}}]{Park2003}%
  \BibitemOpen
  \bibfield  {author} {\bibinfo {author} {\bibfnamefont {S.}~\bibnamefont
  {Park}}, \bibinfo {author} {\bibfnamefont {F.}~\bibnamefont
  {Khalili-Araghi}}, \bibinfo {author} {\bibfnamefont {E.}~\bibnamefont
  {Tajkhorshid}}, \ and\ \bibinfo {author} {\bibfnamefont {K.}~\bibnamefont
  {Schulten}},\ }\href {\doibase 10.1063/1.1590311} {\bibfield  {journal}
  {\bibinfo  {journal} {J. Chem. Phys.}\ }\textbf {\bibinfo {volume} {119}},\
  \bibinfo {pages} {3559} (\bibinfo {year} {2003})}\BibitemShut {NoStop}%
\bibitem [{\citenamefont {Kumar}\ \emph {et~al.}(1992)\citenamefont {Kumar},
  \citenamefont {Rosenberg}, \citenamefont {Bouzida}, \citenamefont
  {Swendsen},\ and\ \citenamefont {Kollman}}]{Kumar1992}%
  \BibitemOpen
  \bibfield  {author} {\bibinfo {author} {\bibfnamefont {S.}~\bibnamefont
  {Kumar}}, \bibinfo {author} {\bibfnamefont {J.~M.}\ \bibnamefont
  {Rosenberg}}, \bibinfo {author} {\bibfnamefont {D.}~\bibnamefont {Bouzida}},
  \bibinfo {author} {\bibfnamefont {R.~H.}\ \bibnamefont {Swendsen}}, \ and\
  \bibinfo {author} {\bibfnamefont {P.~A.}\ \bibnamefont {Kollman}},\ }\href
  {\doibase 10.1002/jcc.540130812} {\bibfield  {journal} {\bibinfo  {journal}
  {J. Comput. Chem.}\ }\textbf {\bibinfo {volume} {13}},\ \bibinfo {pages}
  {1011} (\bibinfo {year} {1992})}\BibitemShut {NoStop}%
\bibitem [{\citenamefont {Roux}(1995)}]{Roux1995}%
  \BibitemOpen
  \bibfield  {author} {\bibinfo {author} {\bibfnamefont {B.}~\bibnamefont
  {Roux}},\ }\href {\doibase http://dx.doi.org/10.1016/0010-4655(95)00053-I}
  {\bibfield  {journal} {\bibinfo  {journal} {Comput. Phys. Commun.}\ }\textbf
  {\bibinfo {volume} {91}},\ \bibinfo {pages} {275} (\bibinfo {year}
  {1995})}\BibitemShut {NoStop}%
\end{thebibliography}

%

\end{document}